\documentclass{jfm}
\usepackage{graphicx}
\usepackage{epstopdf, epsfig}
\usepackage{color}
\usepackage{amssymb}
\usepackage{amsmath}
\usepackage{textcomp} 
\usepackage{hyperref}
\hypersetup{
colorlinks,
citecolor=blue,
linkcolor=blue}

\shorttitle{Transition between advection and propagation in rotating turbulence}
\shortauthor{J.A. Brons, P.J. Thomas and A. Poth\'erat}

\title{Transition between advection and inertial wave propagation in rotating turbulence}

\author{Jonathan A. Brons\aff{1,2}
\corresp{\email{ac2002@coventry.ac.uk}},
  P. J.  Thomas\aff{1}
 \and A. Poth\'erat\aff{2}}

\affiliation{\aff{1}Fluid Dynamics Research Centre, University of Warwick, Coventry, UK
\aff{2}Fluid and Complex Systems Research Centre, Coventry University, Coventry, UK}

\begin{document}

\maketitle

\begin{abstract}
In turbulent flows subject to strong background rotation, the advective mechanisms of turbulence are superseded by the propagation of inertial waves, as the effects of rotation become dominant. While this mechanism has been identified experimentally \citep{dickinson83,davidson06,staplehurst08,kolvin09}, the conditions of the transition between the two mechanisms are less clear. We tackle this question experimentally by tracking the turbulent front away from a solid wall where jets enter an otherwise quiescent fluid. Without background rotation, this apparatus generates a turbulent front whose displacement recovers the  $z(t)\sim t^{1/2}$ law classically obtained with an oscillating grid \citep{dickinson1978_pf} and we further establish the scale-independence of the associated transport mechanism. When the apparatus is rotating at a constant velocity perpendicular to the wall where fluid is injected, not only does the turbulent front become mainly transported by inertial waves, but advection itself is suppressed because of the local deficit of momentum incurred by the propagation of these waves. Scale-by-scale analysis of the displacement of the turbulent front reveals that the transition between advection and propagation is local both in space and spectrally, and takes place when the Rossby number based on the considered scale is of unity, or equivalently, when the scale-dependent group velocity of inertial waves matched the local advection velocity.
\end{abstract}


\section{Introduction}
The main transport mechanism in turbulent flow is advection. When turbulent flows are subject to background rotation, however, inertial waves offer an additional transport mechanism. The competition between them determines the anisotropy and transport properties of rotating turbulence. Here we determine the conditions in which either of them dominates, and especially the scale-dependence of this competition.\\
Turbulence in rotation arises in a variety of industrial and natural contexts, such as centrifuges, precessing spacecrafts or oceanographic and atmospheric flows \citep{vanyo,davidsonA,davidsonB}, where its specific transport and dissipative properties influence or even govern the dynamics of the processes involved. Its most distinctive feature is to form large, more or less, columnar structures aligned with the rotation axis that are perhaps most conspicuous in geophysical flows \citep{pedlosky}. The emergence of columnar structures in rotating flow was first reported in a letter by Kelvin \citep{kelvin1868} and subsequently illustrated in Taylor's famous experiment \citep{taylor1922}. Since then, a number of experiments and numerical simulations have reported the emergence of such columns in turbulent flows \citep{hopfinger1982_jfm,bartello1994_jfm,gallet2015_jfm} and  several scenarii have been proposed to explain their appearance. Underlying the question of how large columnar structures emerge, is that of the processes by which rotating flows and rotating turbulence transport momentum and energy. This question itself hinges on the role played by two essential ingredients of rotating turbulence. The first one is the propagation mechanism associated to linear inertial waves (see \citet{greenspan} for the theory of these waves): for a wavevector $\mathbf k$, with frequency $\omega$ and background rotation $\boldsymbol{\Omega}$, inertial waves follow the dispersion relation, and associated group velocity
\begin{equation}
\omega=\pm2\boldsymbol{\Omega}\cdot \mathbf e_k, \qquad \mathbf v_g=\pm\frac2k \mathbf e_k\times (\boldsymbol{\Omega}\times \mathbf e_k),
\label{eq:iwdisp}
\end{equation}
where $\mathbf e_k=\frac1k\mathbf k$. The preferential transport of momentum along the rotation axis by inertial waves indeed elongates an initially isotropic blob of vorticity along the axis of rotation at a speed of $\Omega t$ \citep{davidson06}, where $\Omega$ is the rotation speed. The second ingredient involves non-linear interactions \citep{cambon1997_jfm, smith99_pof,cambon99_anrev}. In this process, triadic interactions feed an inverse energy cascade towards large scales while non-resonant triads or quartets of waves transfer energy to modes aligned with the axis of rotation. This scenario is supported by numerical simulations and by strong experimental and numerical evidence of an inverse energy cascade \citep{campagne14_pof}. However, Taylor's early experiments in a steady, laminar flow still exhibit anisotropic transport of momentum along the rotation in the absence of waves and non-linearities. This waveless and linear anisotropic transport was indeed recovered in the analytical work of \citet{moore68,moore69}, and \citet{p12_epl}, and suggests that more than a single transport mechanism may exist in rotating flows. Along this line, our recent experiments showed that even in turbulent flows, the anisotropy of the mean flow may not necessarily result from the action of inertial waves or triadic interactions \citep{bpt2019_prl}. Instead, average anisotropy may emerge from an interplay between rotation and non-linear advection, somewhat similar to the interplay between viscous diffusion and rotation in Taylor's laminar flow experiment. Advection and propagation of inertial waves were even found to simultaneously act on fluctuations in nearly two-dimensional flows: while larger scale fluctuations satisfied the dispersion relation for inertial waves, smaller scales behaved as inertial waves \textquotedblleft swept" by the surrounding velocity field of the large quasi-two dimensional structures \citep{campagne15_pre}.\\
With different mechanisms at play, the question arises of their precise respective domain of action, both in terms of the scales concerned and of the main control parameter, the Rossby number $Ro=U/2\Omega l$, that controls the ratio of inertial to Coriolis forces ($U$ and $l$ are typical velocity and lengthscale). One way to tackle the problem experimentally is to track the displacement of a turbulent front when the turbulence is produced by a localised forcing mechanism and progressively invades a domain of an otherwise quiescent fluid. Most experiments of this type involve either jets along the rotation axis or oscillating grids, as respectively pioneered by \citet{mcewans1976_nat} and \citet{dickinson83}. The latter showed that the position of the turbulent front evolved as $z_f\sim t^{1/2}$ as long as the local Rossby number based on $z$ remained greater than unity. Past this point, $Ro$ decreases, turbulence starts to exhibit wave patterns and the front travels as $z_f\sim \Omega t$, as consistent with the group velocity of inertial waves. 

Grid experiments (\citet{staplehurst08}) on the formation of columnar structures in rotating turbulence revealed that around a local critical Rossby number $Ro^{crit}\sim0.4$ the flow transitions from a state where energy and momentum are mostly propagated by inertial waves (below $Ro^{crit}$) to one where they are mostly transported through advection. A recent numerical study on rotating turbulence ignited by a buoyancy anomaly showed that this transition could be spatially localised with some regions dominated by inertial waves, and others where they are absent (\citet{mcdermott2019_jfm}). These authors
also confirmed a critical value of the Rossby number for this transition around 0.5, provided it is built on the correct large scale.

Recent scale-by-scale analysis of the turbulent front further showed that fluctuations were propagated at the group velocity of inertial waves corresponding to their lengthscales, in the limit of strong rotation $Ro\ll1$ (\citet{kolvin09}, turbulence initiated by jets). In statistically steady turbulence, jet experiments  (\citet{yarom14_nat}, $0.006\leq Ro\leq0.2$), and experiments with a 2D mechanical forcing \citep{campagne15_pre} confirmed that some of the fluctuations of frequency lower than $2\Omega$, the maximum frequency of inertial waves, satisfied the dispersion relation for inertial waves (but for the sweeping effect at high wavenumbers identified by \citet{campagne15_pre}). 
The recent experiments of \cite{burmann2018_pf} showed that inertial waves of a wide range of lengthscales emitted by a topography near an Ekman wall could speed up momentum transfer along the rotation axis and lead to an accelerated spin-up time following a step change in the rotation of a cylindrical vessel.\\
Although the role of inertial waves is clearly established in the limit $Ro\ll1$ and in regions of the spectrum where $\omega\leq2\Omega$, the limits of their regime of influence remains unclear, especially in terms of the lengthscales concerned. Both \citet{dickinson83} and \citet{staplehurst08} found that the momentum transport 
mechanism transitions from  propagative regime to an advective one around $Ro$ of the order of unity, however the scale-dependence of this transition remains unexplored.
We set out to examine this question  and, in particular, the scale dependence of the transport mechanisms in a transient turbulent flow under the effect of background rotation. We target regimes where rotation may not dominate over the entire turbulent spectrum. The specific questions we seek to answer are:
\begin{enumerate}
	\item Is there a clear scale separation (in terms of the control parameter and the scales concerned) between advective or non-linear mechanisms on one side, and propagation on the other?
	\item if so, what is the quantitative threshold defining such a separation?
\end{enumerate}
Our approach relies on the tracking of the turbulent front in a flow forced by turbulent jets, with data processing techniques similar to those introduced by \citet{kolvin09} to analyse the scale dependence. The choice of a transient flow presents the advantage that momentum transport can be easily characterised by tracking the progression of the turbulent front. The experimental setup is described in section \ref{sec:experiment}. To characterise the phenomenology of pure advection in our experimental setup, we first analyse non-rotating turbulence in the spirit of \citet{dickinson83} (section \ref{sec:advection}), before running experiments at several rotational velocities (section \ref{sec:propagation}) and drawing conclusions (section \ref{sec:conclusion}).

\section{Experimental methods} 
\label{sec:experiment}
\subsection{Experimental apparatus}
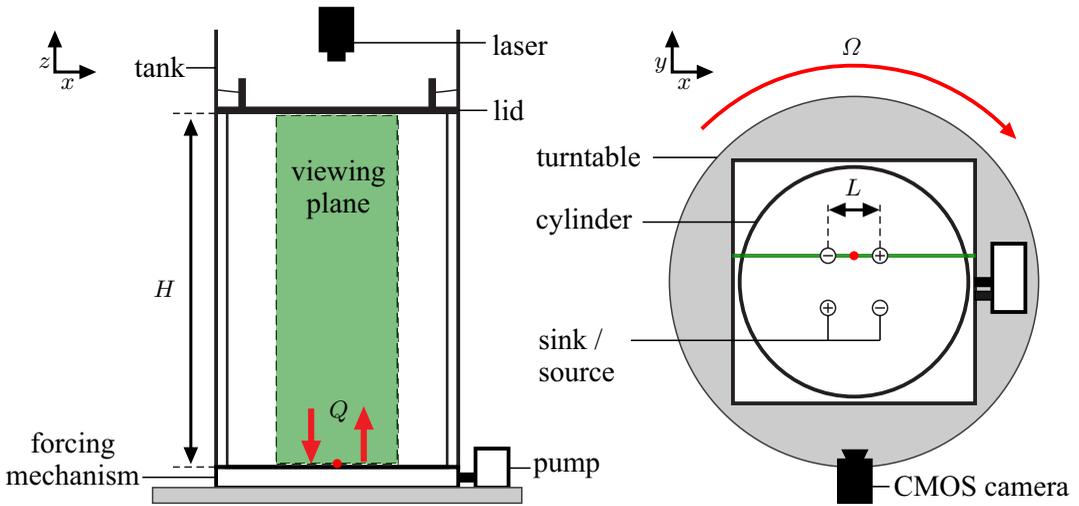
\begin{figure}
	\centerline{
		\includegraphics{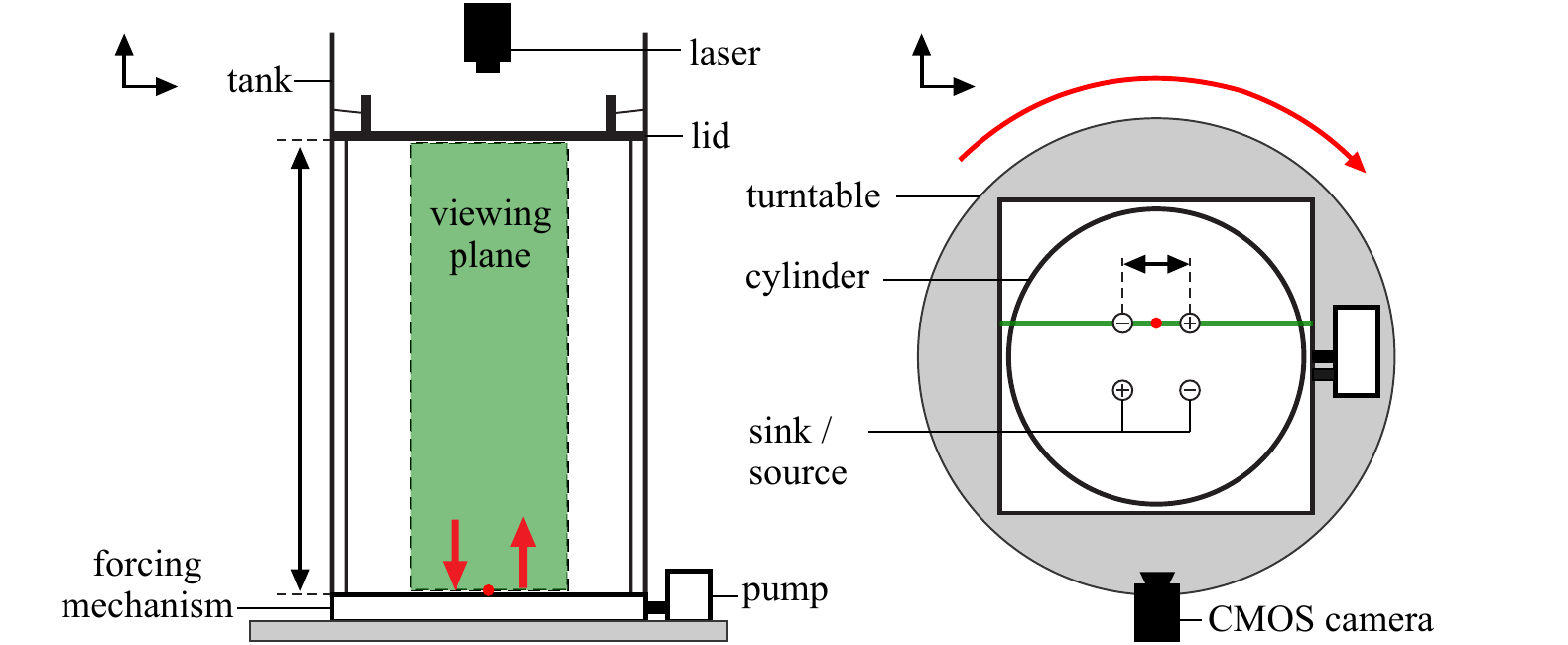}%
		\put(-413,158){\makebox(0,0)[r]{\strut{} $x$}}%
		\put(-422,166){\makebox(0,0)[r]{\strut{} $z$}}%
		\put(-181,158){\makebox(0,0)[r]{\strut{} $x$}}%
		\put(-190,166){\makebox(0,0)[r]{\strut{} $y$}}%
		\put(-375,80){\makebox(0,0)[r]{\strut{} $H$}}%
		\put(-117,118){\makebox(0,0)[r]{\strut{} $L$}}%
		\put(-117,171){\makebox(0,0)[r]{\strut{} $\Omega$}}%
		\put(-310,35){\makebox(0,0)[r]{\strut{} $Q$}}%
	}
	\caption{Sketch of the side- and top-view of the experimental setup, highlighting all important components. The green rectangle shows the approximate size of the flow field recorded and the green line shows the position of the laser sheet across a source/sink pair. Red dots show the position of the origin in our experiments. In top-view (+) refers to a source and (-) to a sink.}
	\label{fig1}
\end{figure}
Figure \ref{fig1} shows a sketch of the setup. The experiment consists of a rectangular tank (60 cm$\times$32 cm$\times$32 cm) centred on a rotating turntable, filled with water (viscosity $\nu=1.0034\times10^{-6}$ m$^2$/s and density $\rho=0.9982\times10^3$ kg/m$^3$). The temperature in the laboratory was kept at 20 \textdegree C.

A forcing mechanism, supported by four pillars at the corners of the mechanism, is placed underneath the bottom wall of the tank. This mechanism forces a flow by injecting and withdrawing fluid through four sources/sinks (diameter $d=1$ mm) located at the corners of a square centred at the bottom wall of the tank. These sources and sinks are respectively identified by the (+) and (-) symbols in figure \ref{fig1}. The distance between the corners of the square is $L=53$ mm. The choice of this square injection pattern provides a quadrupolar flow that remains near the centre of the vessel despite the wide range of Reynolds numbers we investigated. This ensures that measurements made in a fixed region of the flow but at different Reynolds numbers remain comparable to each other. The sources/sinks are connected to an external peristaltic pump via tubing housed underneath the forcing mechanism. The pump (Watson \& Marlow 505-DI) is mounted on the turntable and allows for simultaneous fluid injection through one diagonal of the square (sources) and fluid withdrawal through the other diagonal (sinks), resulting in a zero net mass flux. The forcing mechanism is designed so that the difference in hydraulic resistance across each pair of sources/sinks is kept to a minimum, resulting in a difference in flow rates across these pairs of less than $0.1\%$. The flow rate $Q$ through each of the sources and sinks is considered constant with values of $(0.5,0.9,2.0,3.1,4.7,9.4)\times10^{-6}$ m$^3$/s. A cylinder (height $H$=40 cm, $\varnothing$=30 cm) is placed inside the tank to provide support for a transparent lid placed atop, which prevents surface deformation and gives clear viewing window for the measurement system. \\

During experiments a Coriolis force is applied by spinning the rotating turntable at a constant rotation speed $\Omega$. The turntable is driven by a DC-powered motor connected to the table via a belt-drive. $\Omega$ spanned $\{0,0.52,1.04,2.09,4.19\}$ rad/s with an error on $\Omega$ below $1\%$. The flow field is recorded using a 2D-PIV system. A laser sheet along the $(x,z)$-plane is aligned with a source-sink pair and illuminates an area of approximately 40 cm$\times$ 15 cm, covering the entire height of the tank, as can be seen in figure \ref{fig1}. The laser sheet is generated using a 1 W/532 nm diode-laser and a custom lens system consisting of a concave, a convex and a cylindrical lens. The thickness of the laser sheet remains around 3 mm across the entire height of the flow field. The water is seeded with 10 $\mu$m silver-coated hollow glass spheres, used as tracer particles. Two 1.3MP CMOS cameras are used to record respectively the top and bottom halves of the flow field and cover an area of 21 cm$\times$15 cm each. The recorded areas of these cameras have a small overlapping region of approximately 1 cm at the centre height of the flow field. The cameras record at a frame rate of 60 fps, that is sufficient to resolve the high velocities measured close to the point of fluid injection.\\

For each experimental run, the turntable is initially left to rotate until the fluid inside the tank has reached a state of solid body rotation with rotational
velocity $\Omega$. PIV data are then collected during approximately three seconds to measure the level of residual noise when the liquid is nominally at rest within the rotating frame of reference (\emph{i.e.} in solid body rotation).
Finally the forcing mechanism is activated, at time $t_0=0$, generating a set of jets which penetrate into the flow field. The flow field is recorded for a period of 3 minutes from the time of activation of the forcing mechanism. We identify a time $t_{\rm end}$ for which turbulence occupies the entire vessel. We found $t_{\rm end}<100$ s for all experiments. The injection system is then stopped and the flow is left to decay down to the level of noise recorded in solid body rotation, before the next activation of the injection system.
Velocity fields are derived from recorded images by processing them using the PIVlab software \citep{thielicke} for Matlab. This is done on a 32$\times$32 pixel grid with a 50\% overlap region. The combination of the camera resolution, its field of view and the resolution of the PIV grid result in the smallest resolvable length scale $\ell$=2.1 mm. For each set of experimental parameters, a set of five separate measurements is recorded and the resultant velocity fields are averaged across these separate experiments in order to minimise uncertainties associated to the transient nature of the flow. This method is sufficient to capture the time-dependent event-average of the velocity with a standard deviation of about 5\% across runs. Furthermore, although only the velocity components along the $x$ and $z$ directions are measured, the symmetry of the configuration implies that the flow is statistically invariant by a rotation of $\pm\pi/2$ followed by a reflection about a vertical plane equidistant from two electrodes. As such the 2D measurements provide a good representation of the 3D dynamics, in particular for the purpose of estimating the group velocity of inertial waves of individual horizontal wavenumbers $k$.\\
\subsection{Control parameters}
\begin{table}
	\begin{center}
		\begin{tabular}{lp{0.9cm}p{2.3cm}p{2.6cm}p{1.7cm}p{1.3cm}}
			&Forcing	& $Re_Q$	&$Ek$	&$Ro_Q$	&$z/L$	\\
			\hline
			Currrent	& 4 Jets		&$(0.06-1.2)\times10^4$	&$(4.25-17.0)\times10^{-5}$	&$0.026-2.04$ &$0.1-7.4$ \\
			\\
			\citet{dickinson83}$^*$&	Grid	&n/a		&$\geq4.5\times10^{-6}$		&n/a			& $120-187$\\
			\\
			\citet{staplehurst08}$^*$&	Grid & $83-130$& $(1.44-2.96)\times10^{-6}$	&$0.5-1.4$	& $120-187$\\
			\\
			\citet{kolvin09}$^\dagger$& 248 Jets& $\leq1300$	 &$(6.4-10.8)\times10^{-6}$	&$\leq0.021$	&$1.4-11.4$\\
		\end{tabular}
		\caption{Comparison between the parameter range explored in the current experiment and experiments conducted in other studies. $^*$Ekman numbers are based on the containers heights. $^\dagger$Based on $\Omega$ and upper bound for $Q$ given by Kolvin \etal.}
		\label{tab1}
	\end{center}
\end{table}
We chose a rotating frame of reference with origin centred between two adjacent corners of the square, $\mathbf e_x$ and $\mathbf e_y$ in the horizontal plane and $\mathbf e_z$ pointing upwards, indicated by the red dot in figure \ref{fig1}.\\
Both $Q$ and $\Omega$ provide control over two non-dimensional governing parameters, namely the Ekman number $Ek = \nu/2\Omega L^2 \in[17.0, 8.50, 4.25]\times10^{-5}$ and a Reynolds number based on the flow rate, $Re_Q = U_0d/\nu \in[600,1200,2500,4000, 6000, 12000]$, where $U_0=4Q/\pi d^2$. Here $L$ is chosen as the characteristic lengthscale to make comparison easier to previous experiments. Results are presented in non-dimensional form, using $L$ and $U_0$ as reference length and velocity scales respectively. In comparison to the current experiment, \citep{kolvin09} applied a significantly stronger Coriolis force, while inertial forces were almost always weaker. This difference in parameters reflects a difference in purpose between both setups: while \citet{kolvin09}'s work targeted the limits of high rotation, and low inertia, we are targeting a transitional regime where inertia and the Coriolis force compete. Their ratio is measured by a Rossby number based on the injected velocity $Ro_Q=EkRe_Q$. For comparison with previous experiments on rotating turbulent fronts, the attainable values of the non-dimensional parameters are reported in table \ref{tab1}.

\subsection{Data analysis}
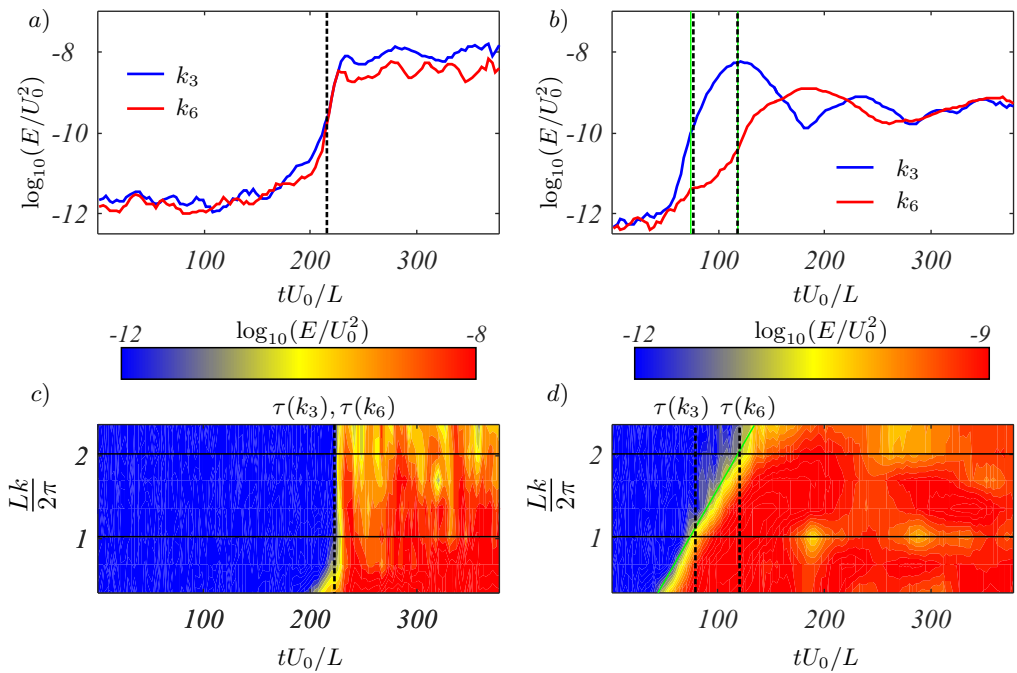
\begin{figure}
	\centerline{
		\includegraphics{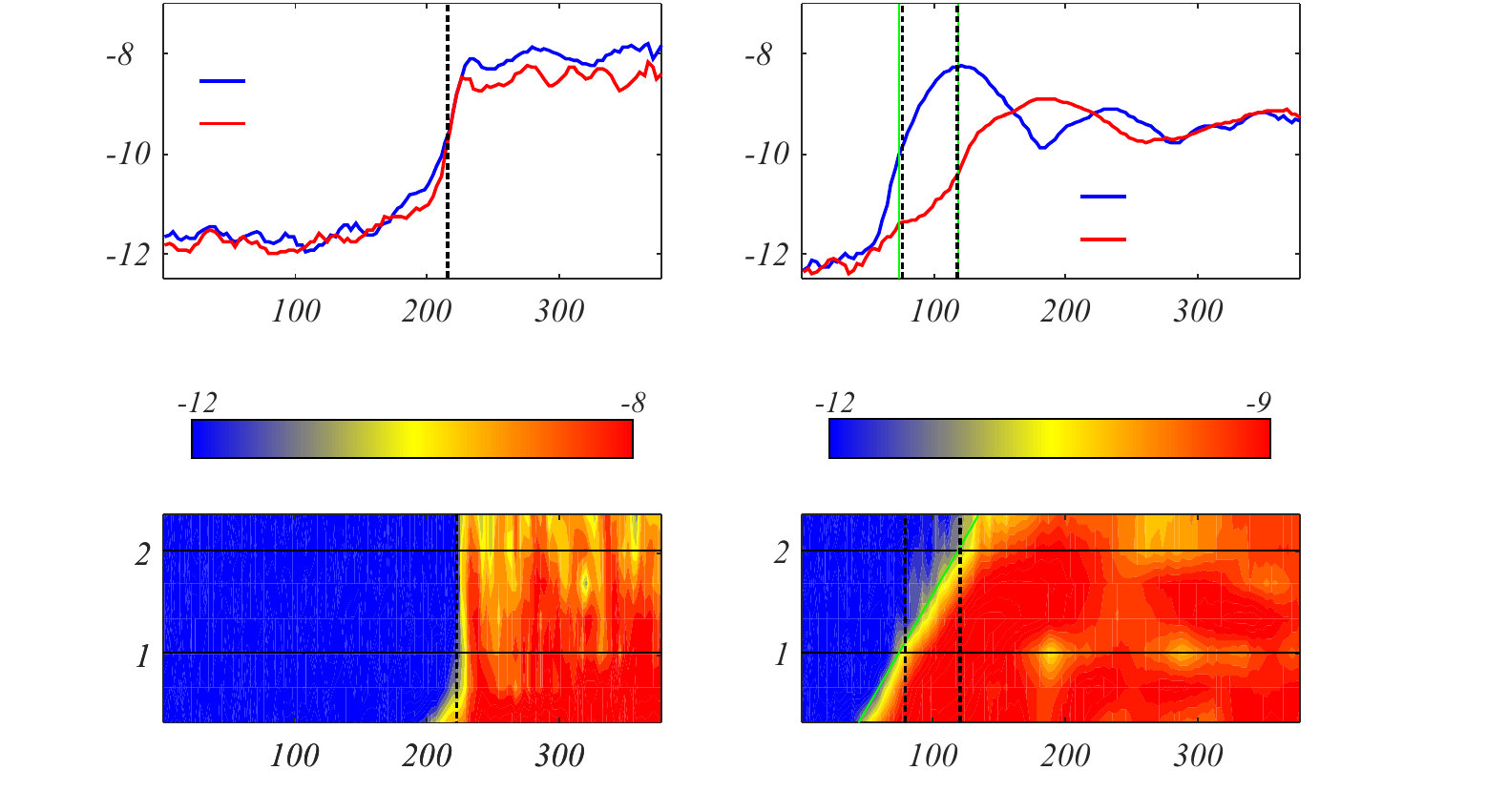}
		\put(-425,241){\makebox(0,0)[r]{\strut{} $a)$}}%
		\put(-230,241){\makebox(0,0)[r]{\strut{} $b)$}}%
		\put(-425,101){\makebox(0,0)[r]{\strut{} $c)$}}%
		\put(-230,101){\makebox(0,0)[r]{\strut{} $d)$}}%
		\put(-98,185){\makebox(0,0)[r]{\strut{} $k_3$}}%
		\put(-98,173){\makebox(0,0)[r]{\strut{} $k_6$}}%
		\put(-369,220){\makebox(0,0)[r]{\strut{} $k_3$}}%
		\put(-369,208){\makebox(0,0)[r]{\strut{} $k_6$}}%
		\put(-315,138){\makebox(0,0)[r]{\strut{} $tU_0/L$}}%
		\put(-121,138){\makebox(0,0)[r]{\strut{} $tU_0/L$}}%
		\put(-315,2){\makebox(0,0)[r]{\strut{} $tU_0/L$}}%
		\put(-121,2){\makebox(0,0)[r]{\strut{} $tU_0/L$}}%
		\put(-295,96){\makebox(0,0)[r]{\strut{} $\tau(k_3),\tau(k_6)$}}%
		\put(-177,96){\makebox(0,0)[r]{\strut{} $\tau(k_3)$}}%
		\put(-152,96){\makebox(0,0)[r]{\strut{} $\tau(k_6)$}}%
		\put(-432,71){\rotatebox{-270}{\makebox(0,0)[r]{\strut{}\Large $\frac{Lk}{2\pi}$}}}%
		\put(-432,221){\rotatebox{-270}{\makebox(0,0)[r]{\strut{} $\log_{10}(E/U_0^2)$}}}%
		\put(-237,71){\rotatebox{-270}{\makebox(0,0)[r]{\strut{}\Large $\frac{Lk}{2\pi}$}}}%
		\put(-237,221){\rotatebox{-270}{\makebox(0,0)[r]{\strut{} $\log_{10}(E/U_0^2)$}}}%
		\put(-110,125){\rotatebox{0}{\makebox(0,0)[r]{\strut{} $\log_{10}(E/U_0^2)$}}}%
		\put(-305,125){\rotatebox{0}{\makebox(0,0)[r]{\strut{} $\log_{10}(E/U_0^2)$}}}%
	}%
	\caption{a,b) Temporal energy profiles $E(k,t)$ for modes $k_{3}L/2\pi\approx1.0$ and $k_{6}L/2\pi\approx2.0$ at $Re_Q=2500$ and a height $z/L = 4.91$. c,d) Contour plots of $E(k,t)$, where solid black lines highlight $E(k,t)$ for modes $k_3$ and $k_6$. Experiments conducted at a,c) $Ek=\infty$ and b,d) $Ek=4.25\times10^{-5}$. Arrival times ($\tau(k_3),\tau(k_6)$) are represented by dashed lines. Green lines in b) represent the theoretical arrival time for inertial waves of wavenumber $k_3,k_6$ respectively. Similarly, the green line in d) represent the theoretical contour of linear inertial wave propagation across the entire spectrum.}
	\label{fig2}
\end{figure}

To differentiate advective from propagative processes, we shall analyse the scale dependence of the evolution of the turbulent front. For this, we follow a method similar to \citet{kolvin09}: we first apply a discrete Fourier transform along $x$ to the velocity field $\mathbf{u}(x,z,t)$ to obtain a space and time-dependent power density spectrum $E(k,z)=|\hat{\mathbf{u}}(k,z,t)^2|$, expressed in term of wavenumber $k$. This operation is performed for each acquisition timestep $t$. From this, variations of energy at one spatial location for a given wavelength are extracted by fixing $z$ and $k$.\\
Figure \ref{fig2} (a) and (b) shows example representations of $E(k,z,t)$ at $z/L=4.91$ for one non-rotating and one rotating experiment, respectively. In each case, The time-variations of $E(k,z,t)$ exhibit a sharp transition from an initially low energy state at noise level to a high energy, turbulent state. For any mode $k$, we consider that the front has arrived at height $z$ at arrival time $\tau$ for the lowest value $\tau$ of t such that $E(k,z,t)$ exceeds a threshold value between these two states.
For each set of parameters $(Re_Q,Ek)$ and each value of $k$, the threshold value $E_T(k)$ is defined as $E_T(k)=\frac1{2H}\int_0^H E(k,z,t_0)dz+\frac1{2H}\int_0^H E(k,z,t_{\rm end})dz$, \emph{i.e.}, the average between the state of residual noise at $t<t_0$, and the state when turbulence has invaded the full domain at $t=t_{\rm end}$. The time of arrival at a prescribed height $z$ of a given modes $k$ is obtained as the time $\tau$ such that $E(k,z,\tau)=E_T(k)$. The position of the front at time $\tau$ of the physical domain containing energy in mode $k$ is then simply tracked through the location $z(\tau)$ for which $E(k,z,\tau)=E_T(k)$. Additionally, the evolution of the spectral shape of the turbulent cloud is visualized by plotting contours of $E(k,z,t)$ as examplified in figure \ref{fig2} (c) and (d) for the same two experimental cases.\\
Figure \ref{fig2} (a) shows that mode $k_3$ and $k_6$ display the same variations in energy at all times, with both modes arriving at roughly the same time $\tau_3U_0/L\approx\tau_6U_0/L\approx215$. This is reflected in the near vertical contour in figure \ref{fig2} (c). Figure \ref{fig2} (b) however shows that mode $k_3$ progresses substantially faster than mode $k_6$ arriving at time $\tau_3U_0/L\approx80$ and $\tau_6U_0/L\approx115$, respectively. This difference in displacement velocity observed in the rotating case translates into the slanted contour of figure \ref{fig2} (d).\\
\section{Advection of the turbulent cloud with and without background rotation}
\label{sec:advection}

\subsection{Non-rotating jet experiments}
\begin{figure}
	\centerline{
		\includegraphics{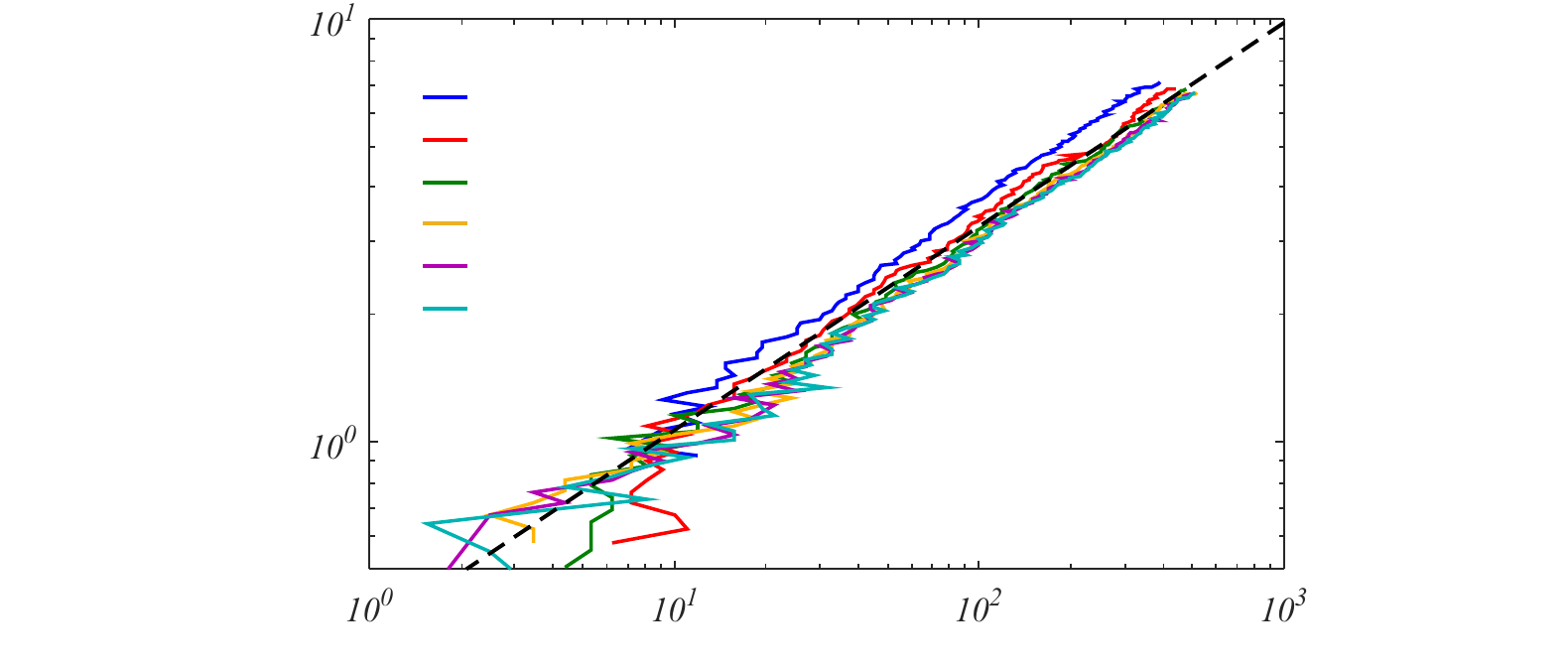}
		\put(-303,165){\makebox(0,0)[r]{\strut{} $k_{1}$}}%
		\put(-303,152){\makebox(0,0)[r]{\strut{} $k_{2}$}}%
		\put(-303,140){\makebox(0,0)[r]{\strut{} $k_{3}$}}%
		\put(-303,128){\makebox(0,0)[r]{\strut{} $k_{4}$}}%
		\put(-303,116){\makebox(0,0)[r]{\strut{} $k_{5}$}}%
		\put(-303,103){\makebox(0,0)[r]{\strut{} $k_{6}$}}%
		\put(-120,100){\makebox(0,0)[r]{\strut{} $0.351~ \left(\frac{\tau U_0}{L}\right)^{0.482}$}}%
		\put(-375,125){\rotatebox{-270}{\makebox(0,0)[r]{\strut{} $(z(\tau)-z_0)/L$}}}%
		\put(-200,5){\makebox(0,0)[r]{\strut{} $\tau U_0/L$}}%
	}
	\caption{Arrival time $\tau$ at height $z$ for the first six modes $k_{i}$ at $Re_Q=6000$ in the absence of rotation ($Ek=\infty$). The dashed line is a fit of the experimental data for $z\geq0.8$.}
	\label{fig3}
\end{figure}

We first analyse the motion of the turbulent front in the absence of a Coriolis force (\emph{i.e.} $Ek=\infty$), where no propagative behaviour is expected, to enable us to quantify changes in behaviour when rotation is present. Under these circumstances the only available mechanism is advection. Figure \ref{fig3} shows  the motion of the front at $Re_Q=6000$ for the first six modes of the Discrete Fourier Transform with wavenumber $k_i$, where $\{k_{i}\}_{i=1..6}=2\pi i/(N\ell)$. Here $N$ is the number of PIV grid-points along the horizontal plane ($N$=64) and $N\ell$ is this width of the resolved horizontal plane which is about 155 mm.
For $Ek=\infty$ the motion of the turbulent front is independent of $k$. The position of the turbulent front follows a scaling of  $(z-z_0)/L\approx (0.351\pm 0.016) {(\tau U_0/L)}^{0.482\pm 0.011}$ across all scales of the flow. This behaviour is observed for all $Re_Q$ explored. Here, the offset $z_0$ is calculated so that the power law fit extends to $\tau=0$.
\begin{figure}
	\centerline{
		\includegraphics{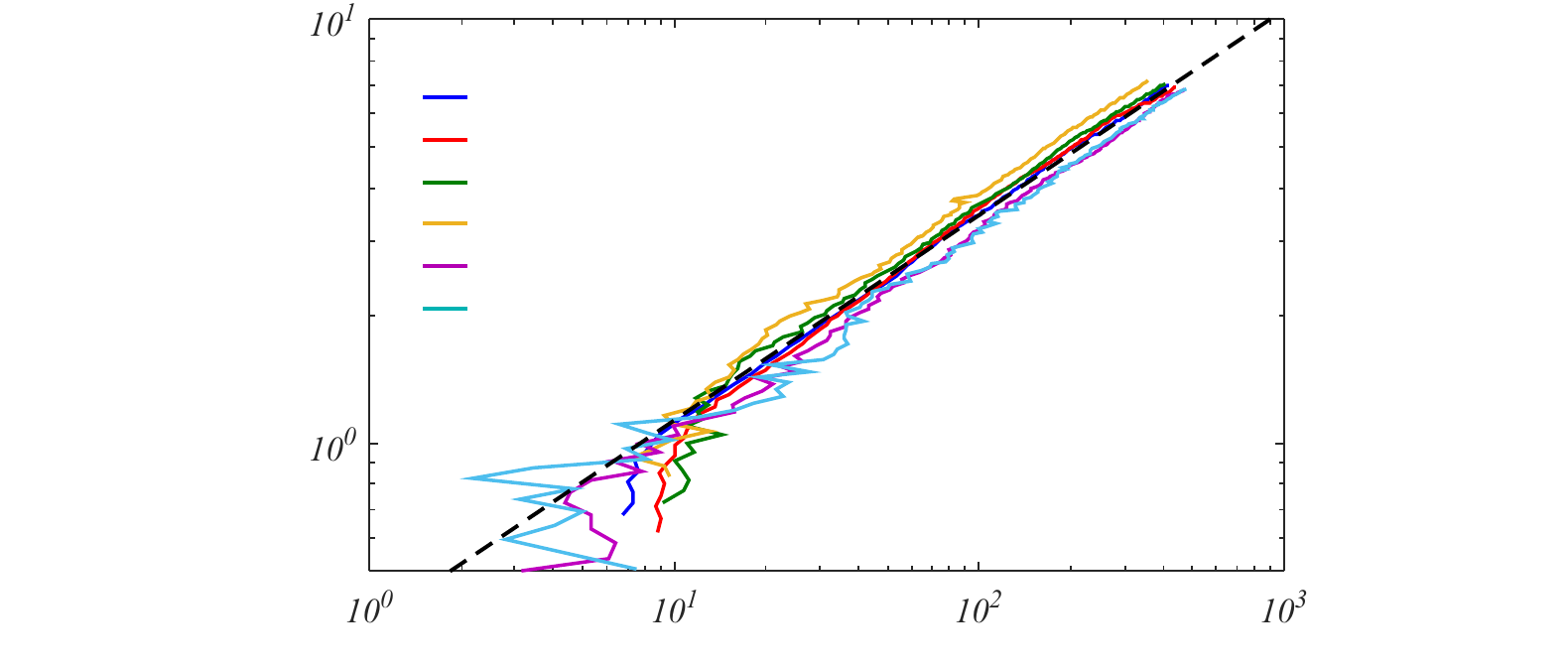}
		\put(-292,177){\makebox(0,0)[r]{\strut{} $Re_Q$}}%
		\put(-294,164){\makebox(0,0)[r]{\strut{} $600$}}%
		\put(-292,152){\makebox(0,0)[r]{\strut{} $1200$}}%
		\put(-292,140){\makebox(0,0)[r]{\strut{} $2500$}}%
		\put(-292,128){\makebox(0,0)[r]{\strut{} $4000$}}%
		\put(-292,116){\makebox(0,0)[r]{\strut{} $6000$}}%
		\put(-290,104){\makebox(0,0)[r]{\strut{} $12000$}}%
		\put(-120,100){\makebox(0,0)[r]{\strut{} $0.377~ {\left(\frac{\tau U_0}{L}\right)}^{0.483}$}}%
		\put(-375,125){\rotatebox{-270}{\makebox(0,0)[r]{\strut{} $({\bar z(\tau)}-\bar{z_0})/L$}}}%
		\put(-200,5){\makebox(0,0)[r]{\strut{} $\tau U_0/L$}}%
	}
	\caption{Arrival time $\tau$ at height $z$ at $Ek=\infty$ across all $Re_Q$, where $\tau$ is taken as the average across first six modes $k_{i}$. The solid black line is a fit of data where $z/L\geq0.8$}
	\label{fig4}
\end{figure}
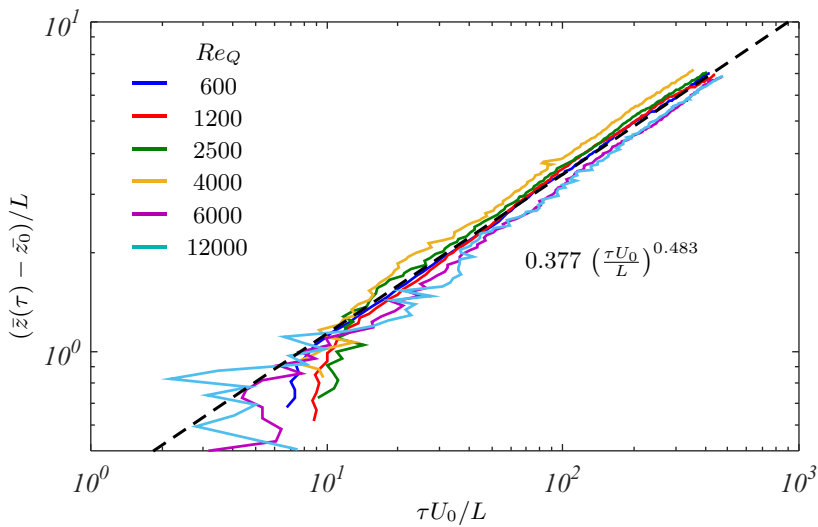
Since the jet is turbulent, and that all scales are displaced at the same velocity, the position of the turbulent front $\bar z(\tau)$ may be calculated as the average over the first six modes of the Discrete Fourier Transform used to calculate $E(k,z,t)$. Figure \ref{fig4} shows the variations of $\bar z(\tau)$ with $Re_Q$ in the absence of Coriolis force. By non-dimensionalizing $\tau$ by the characteristic injection time $L/U_0$ the data for $\bar z$ collapses almost onto a single line for $\bar z/L\geq0.8$. This shows that in the absence of rotation the non-dimensional arrival time is determined solely by the injection velocity $U_0$ as
\begin{equation}
\frac{\bar z(\tau)-\bar z_0}{L}= (0.377\pm 0.014) \times \left(\frac{\tau U_0}{L}\right)^{0.483\pm 0.010},
\label{eq_Z_no_rot}
\end{equation}
with corresponding velocity of the turbulent front as it progresses in the quiescent fluid,
\begin{equation}
\frac{U(z)}{U_0}= ~(6.41\pm 0.11)\times 10^{-2} \left(\frac{z}{L}\right)^{-1.070\pm 0.027}~.
\label{eq_Uz_adv}
\end{equation}
The $z^{-1}$-profile closely resembles the axial velocity profile of a single steady turbulent jet \citep[p.100]{pope}, most likely because of the nature of our forcing. Nevertheless, the fact that the transient jet exhibits the same profile as the statistically steady jet, indicates that the jet develops in such a way that the flow behind the front is in a statistically steady state even though the front continues to progress. In other words, the front \textquotedblleft sweeps" through the quiescent fluid, leaving a statistically steady turbulent flow behind. This is confirmed by observing the shape of the turbulent region as seen figure \ref{fig5}(a). This region only evolves by extending upwards, as the front progresses, but not radially. This scenario is further supported as for the profile of a steady jet, momentum conservation implies that the turbulent region should grow linearly with the distance to the origin as a result. Figure \ref{fig5}(a) seems to confirm that the jet behind the turbulent front satisfies this property.\\
\\
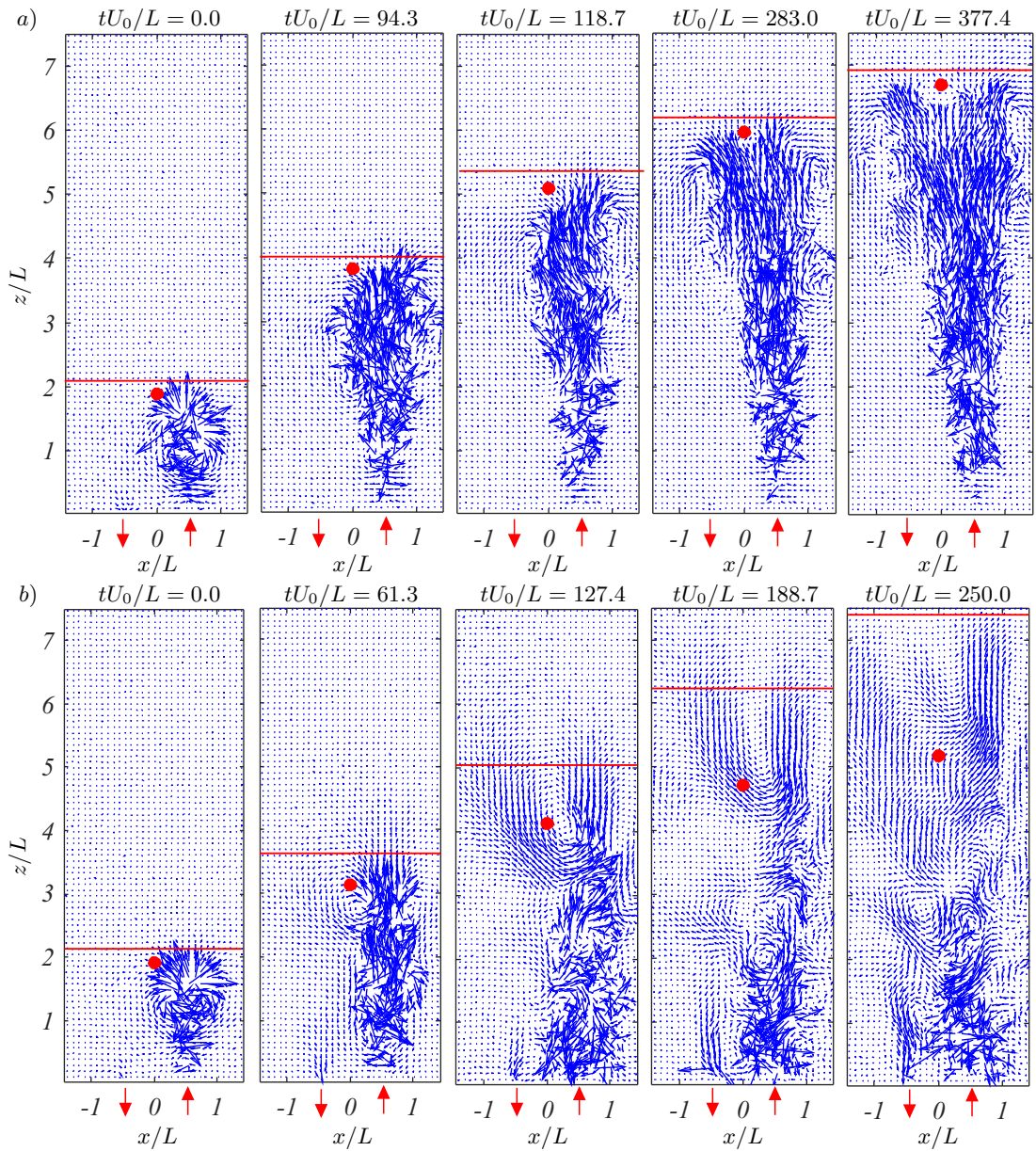
\begin{figure}
	\centerline{
		\includegraphics{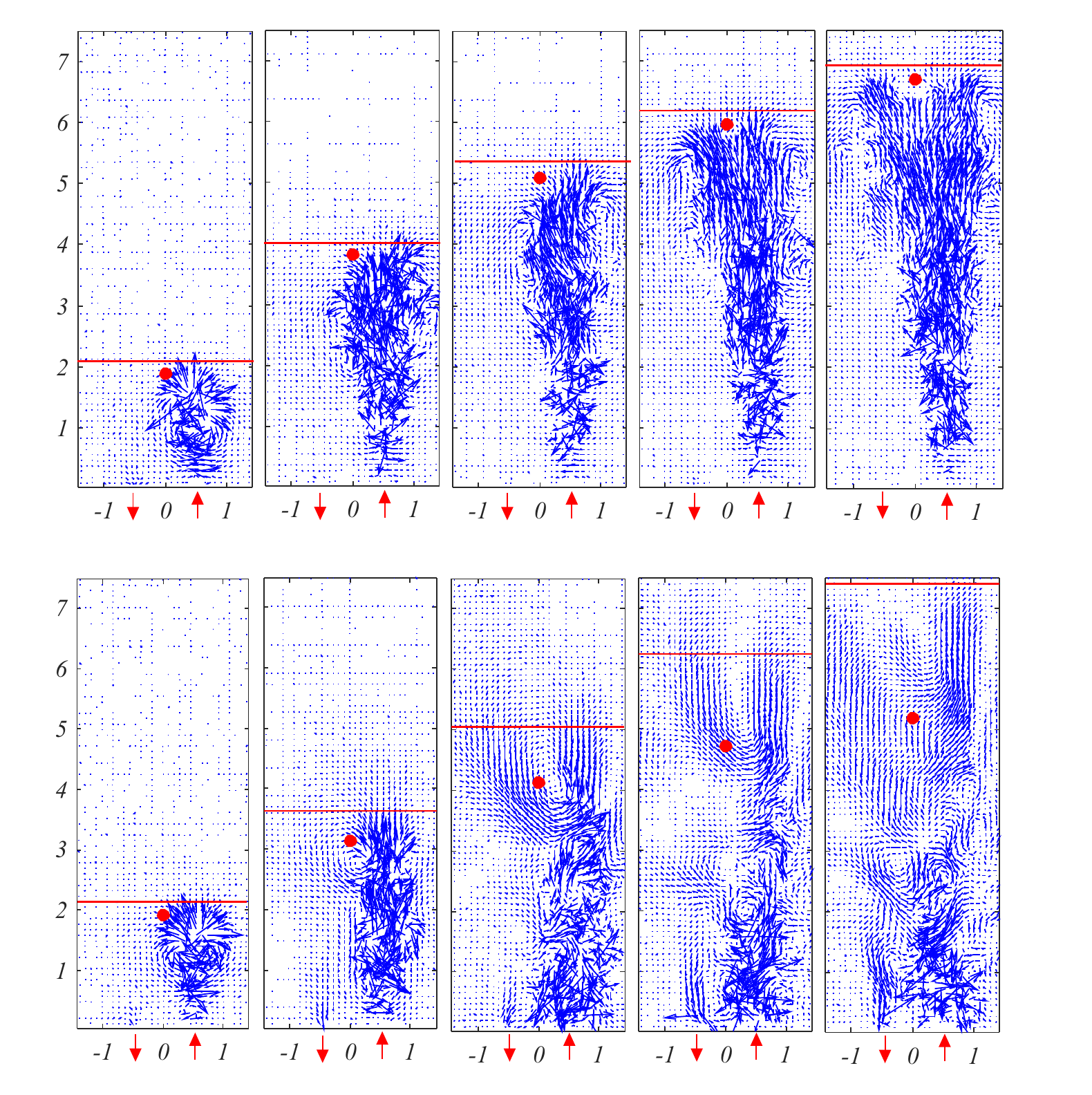}%
		\put(-67,5){\makebox(0,0)[r]{\strut{} $x/L$}}%
		\put(-145,5){\makebox(0,0)[r]{\strut{} $x/L$}}%
		\put(-223,5){\makebox(0,0)[r]{\strut{} $x/L$}}%
		\put(-301,5){\makebox(0,0)[r]{\strut{} $x/L$}}%
		\put(-379,5){\makebox(0,0)[r]{\strut{} $x/L$}}%
		\put(-46,221){\makebox(0,0)[r]{\strut{} $tU_0/L=250.0$}}%
		\put(-123,221){\makebox(0,0)[r]{\strut{} $tU_0/L=188.7$}}%
		\put(-200,221){\makebox(0,0)[r]{\strut{} $tU_0/L=127.4$}}%
		\put(-283,221){\makebox(0,0)[r]{\strut{} $tU_0/L=61.3$}}%
		\put(-362,221){\makebox(0,0)[r]{\strut{} $tU_0/L=0.0$}}%
		\put(-67,233){\makebox(0,0)[r]{\strut{} $x/L$}}%
		\put(-145,233){\makebox(0,0)[r]{\strut{} $x/L$}}%
		\put(-223,233){\makebox(0,0)[r]{\strut{} $x/L$}}%
		\put(-301,233){\makebox(0,0)[r]{\strut{} $x/L$}}%
		\put(-379,233){\makebox(0,0)[r]{\strut{} $x/L$}}%
		\put(-46,450){\makebox(0,0)[r]{\strut{} $tU_0/L=377.4$}}%
		\put(-123,450){\makebox(0,0)[r]{\strut{} $tU_0/L=283.0$}}%
		\put(-200,450){\makebox(0,0)[r]{\strut{} $tU_0/L=118.7$}}%
		\put(-283,450){\makebox(0,0)[r]{\strut{} $tU_0/L=94.3$}}%
		\put(-362,450){\makebox(0,0)[r]{\strut{} $tU_0/L=0.0$}}%
		\put(-442,125){\rotatebox{-270}{\makebox(0,0)[r]{\strut{} $z/L$}}}%
		\put(-435,221){\makebox(0,0)[r]{\strut{} $b)$}}%
		\put(-442,352){\rotatebox{-270}{\makebox(0,0)[r]{\strut{} $z/L$}}}%
		\put(-435,450){\makebox(0,0)[r]{\strut{} $a)$}}%
	}
	\caption{Snapshots of the jet velocity field for $Re_Q=2500$ and a) $Ek=\infty$ and b) $Ek=8.50\times10^{-5}$. The red dot shows the position $z^a(t)$ of a numerical particle initially positioned at $z_0/L=2$, where $t=0$ coincides $z/L=2$. The red line shows the position of the front. The small difference in position between particle and front in (a) is artificial and caused by differences in sensitivity in the methods used to measure their position. Red arrows indicate the point of fluid injection/withdrawal. Supplementary material: \textit{movie1.avi} shows the simultaneous evolution of both jets represented here.}
	\label{fig5}
\end{figure}
Scaling (\ref{eq_Z_no_rot}) is near-identical to the front displacement law found experimentally by \citet{dickinson1978_pf} with an oscillating grid instead of jets. This law is itself in agreement with the theoretical prediction of \citet{long1978_pf}, expressed dimensionally as $z_{\rm dim}(t)\sim K t_{\rm dim}^{1/2}$, where constant $K$ is expected to scale with the action generating the turbulence. While an exact determination of the parameters governing the variation of this quantity is not available in \citet{dickinson1978_pf}'s grid experiments, $K\simeq (0.43\pm 0.02) (U_0L)^{1/2}$ in the present case of jet-driven turbulence. The displacement offset $\bar z_{0 \rm dim}$ lies in the range 0.5-2.0 cm, similar to the experiments of \citet{dickinson1978_pf} and \citet{hopfinger1976_jfm}, most likely on the grounds that the small scale forcing from the grid and the jets lie in the same range of scales. Additionally, $z_0$ exhibits no variations of significance with either the wavelength considered or $Re_Q$ (see figure \ref{fig6}, beyond fluctuations within the measurement error, which we estimate to approximately 0.5 cm). These results confirm that the 4-jet system generates a turbulent front  with the same dynamics as the classic oscillating grid. Moreover they establish the scale-independence of the advective front motion.\\
Physically, $z_0$ corresponds the virtual point from where turbulent advection starts. A possible reason for $z_0$ not to be zero is that the jet is not turbulent at $\tau=0$: the first state of the development of the jet is laminar, followed by the development of instabilities, which in turn lead to turbulence in a finite time. Hence the initial advection may not follow the turbulent advection law. It follows that if the advected position of the turbulent front is extrapolated back to $\tau=0$ according to that law, the result may not coincide with the bottom of the tank but with an offset position $z_0$.\\  
\begin{figure}
	\centerline{
		\includegraphics{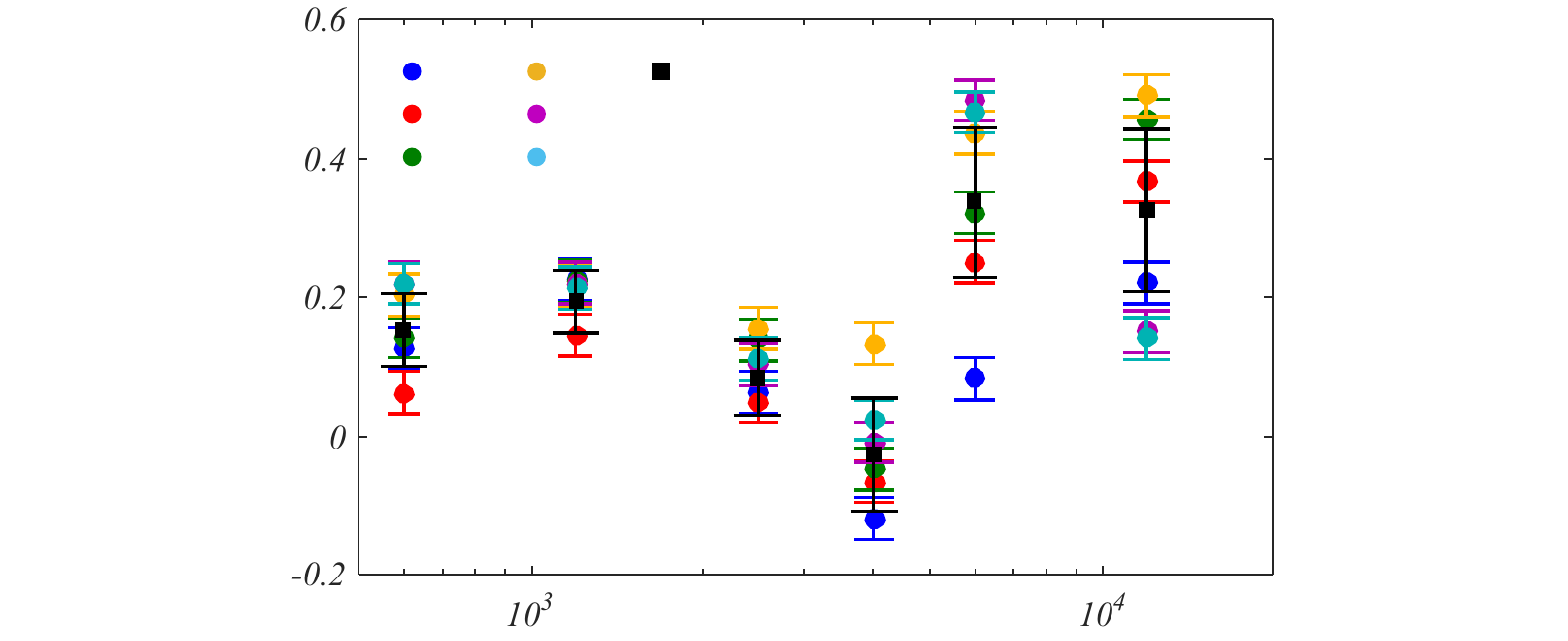}
		\put(-320,165){\makebox(0,0)[r]{\strut{} $k_1$}}%
		\put(-320,152){\makebox(0,0)[r]{\strut{} $k_2$}}%
		\put(-320,140){\makebox(0,0)[r]{\strut{} $k_3$}}%
		\put(-283,165){\makebox(0,0)[r]{\strut{} $k_4$}}%
		\put(-283,152){\makebox(0,0)[r]{\strut{} $k_5$}}%
		\put(-283,140){\makebox(0,0)[r]{\strut{} $k_6$}}%
		\put(-238,165){\makebox(0,0)[r]{\strut{} $\bar{z_0}/L$}}%
		\put(-379,115){\rotatebox{-270}{\makebox(0,0)[r]{\strut{} $z_0/L$}}}%
		\put(-200,3){\makebox(0,0)[r]{\strut{} $Re_Q$}}%
	}
	\caption{Offset $z_0$ measured across $Re_Q$ and scales $k_i$ at $Ek=\infty$.}
	\label{fig6}
\end{figure}
\subsection{Advection in the presence of background rotation}
\begin{figure}
	\centerline{
		\includegraphics{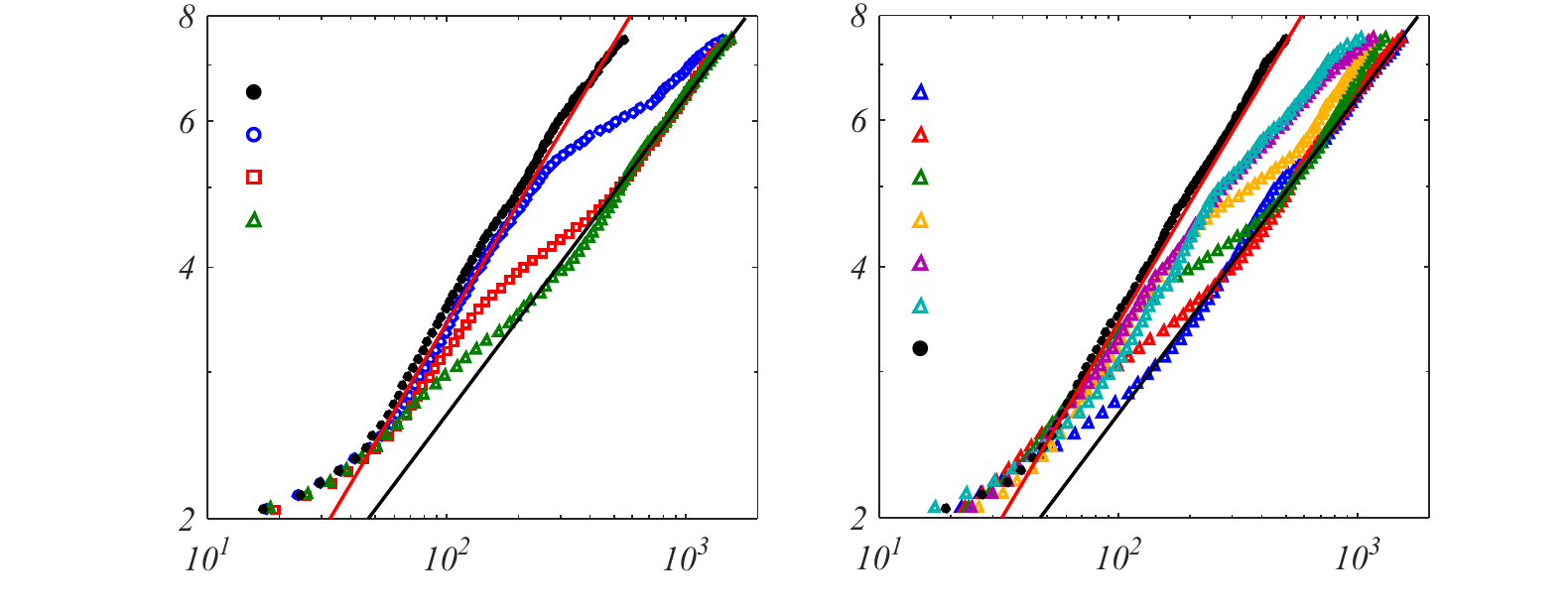}
		\put(-220,174){\makebox(0,0)[r]{\strut{} $b)$}}%
		\put(-110,4){\makebox(0,0)[r]{\strut{} $tU_0/L$}}%
		\put(-220,120){\rotatebox{-270}{\makebox(0,0)[r]{\strut{} $z^a(t)/L$}}}%
		\put(-155,165){\makebox(0,0)[r]{\strut{} $Re_Q$}}%
		\put(-156,152){\makebox(0,0)[r]{\strut{} $600$}}%
		\put(-154,139){\makebox(0,0)[r]{\strut{} $1200$}}%
		\put(-154,127){\makebox(0,0)[r]{\strut{} $2500$}}%
		\put(-154,114){\makebox(0,0)[r]{\strut{} $4000$}}%
		\put(-154,101){\makebox(0,0)[r]{\strut{} $6000$}}%
		\put(-152,89){\makebox(0,0)[r]{\strut{} $12000$}}%
		\put(-152,75){\makebox(0,0)[r]{\strut{} $Ek~\infty$}}%
		\put(-76,43){\makebox(0,0)[r]{\strut{} $0.48~\left(\frac{tU_0}{L}\right)^{0.38}$}}%
		\put(-85,165){\makebox(0,0)[r]{\strut{}\textcolor{red}{$0.38~\left(\frac{tU_0}{L}\right)^{0.48}$}}}%
		\put(-413,174){\makebox(0,0)[r]{\strut{} $a)$}}%
		\put(-305,4){\makebox(0,0)[r]{\strut{} $tU_0/L$}}%
		\put(-413,120){\rotatebox{-270}{\makebox(0,0)[r]{\strut{} $z^a(t)/L$}}}%
		\put(-350,164){\makebox(0,0)[r]{\strut{} \small $Ek$}}%
		\put(-350,151.5){\makebox(0,0)[r]{\strut{} \small $\infty$}}%
		\put(-325,139){\makebox(0,0)[r]{\strut{} \small $17.0\times10^{-5}$}}%
		\put(-325,126.5){\makebox(0,0)[r]{\strut{} \small $8.50\times10^{-5}$}}%
		\put(-325,114){\makebox(0,0)[r]{\strut{} \small $4.25\times10^{-5}$}}%
		\put(-275,43){\makebox(0,0)[r]{\strut{}$0.48~\left(\frac{tU_0}{L}\right)^{0.38}$}}%
		\put(-335,80){\makebox(0,0)[r]{\strut{}\textcolor{red}{$0.38~\left(\frac{tU_0}{L}\right)^{0.48}$}}}%
	}%
	\caption{Position $z^a$ of a particle initially placed at $z_0/L=2$ as function of time. (a) $Re_Q=1200$ with varying $Ek$. (b) $Ek=4.25\times10^{-5}$ with varying $Re_Q$. Supplementary material: \textit{movie2.avi} contains a video showing the evolution of the jet next to the evolution of $z^a(t)$ and $z(t)$. Time $t=0$ corresponds to the time when the particle is first displaced}.
	\label{fig7}
\end{figure}

As mentioned in introduction, one of the main reasons for the choice the transient jet configuration is that momentum transport can easily be tracked through the displacement of the turbulent front. In order to differentiate advection from other momentum transport mechanisms in the rotating jet, we first need to understand how rotation affects advection itself. This is done by calculating the Lagrangian flow $\Phi$ associated to the two-dimensional flow field obtained from the PIV measurements for $\mathbf u(x,z,t)$. For a particle initially located at $\mathbf r_0=(x(t=0),z(t=0))=\mathbf r(t=0)$,
\begin{equation}
\mathbf r(t)=\Phi(\mathbf r_0,t)=\int_0^t \mathbf u(\mathbf r(t'),t')dt'.
\end{equation}
For the purpose of determining the motion of the turbulent front, we shall consider advection of a particle in the $z$ direction only and calculate its virtual motion if it were purely advected by the jet. Additionally, since we are interested in the movement of the front and not of an actual particle, we shall consider the maximum advection velocity across the $x$ direction rather than the local one and define the purely advective displacement as
\begin{equation}
z^a(t)=\int_0^t \max_{x}\{\mathbf u(x,z^a(t'),t')\cdot \mathbf e_z\}dt'.
\end{equation}
It is noteworthy that the coordinate $z^a(t)$ does not track an actual fluid particle. 
Indeed, while fluid transport indeed occurs through advection, it does not occur through propagation of inertial waves. Momentum transport, on the other hand, does occur in both processes. More precisely, the turbulent front materialises the transport of the fluctuating part of the momentum.
The evolution of $z^a(t)$ is represented in two ways: figure \ref{fig7} (a) shows $z^a(t)$ for $Re_Q=1200$ and varying $Ek$, while \ref{fig7} (b) shows $z^a(t)$ for $Ek=4.25\times10^{-5}$ and varying $Re_Q$. Here $t=0$ is set to the time when the particle is first displaced from its initial location 
at $z^a(0)/L\approx2$. Figure \ref{fig5} show snapshots of the jet velocity field with and without rotation, with the position $z^a$ represented by a single particle. In the absence of rotation the position of the particle closely follows that of the turbulent front \emph{i.e.} $z^a(t)\approx z(t)$. When rotation is present, the advected particles initially follows the turbulent front but falls well behind after this initial phase. The beginning of this second phase, which can be identified in figure \ref{fig7} as the point where the curves deviate from the $Ek=\infty$ case, coincides with the appearance of chevron-patterns in the velocity field. These patterns are visible in figure \ref{fig5} for $tU_0/L\geq 127.4$, and in the supplementary material: \textit{movie1.avi}.
They are a signature of inertial waves being emitted by the jets. A combination of frequency filtering and phase averaging \citep{cortet10_pof} revealed that the jets emit inertial wave packets of all possible frequencies $\omega<2\Omega$ and propagation angles $\theta=\arccos(e_\mathbf k\cdot e_\mathbf z)$ corresponding to the dispersion relation of inertial waves (\ref{eq:iwdisp}). The chevron patterns are a superposition of numerous waves, which allows us to visualise the inertial waves. The details of this frequency analysis are reported in \citet{bpt2019_prl}.

The slowdown of advection can be understood in terms of momentum conservation: since part of the momentum is conveyed by inertial waves ahead of the \textquotedblleft purely advected" position, less momentum is locally available for purely advective momentum transport. The effect is all the more visible as rotation is important.\\

\begin{figure}
	\centerline{
		\includegraphics[scale=0.9]{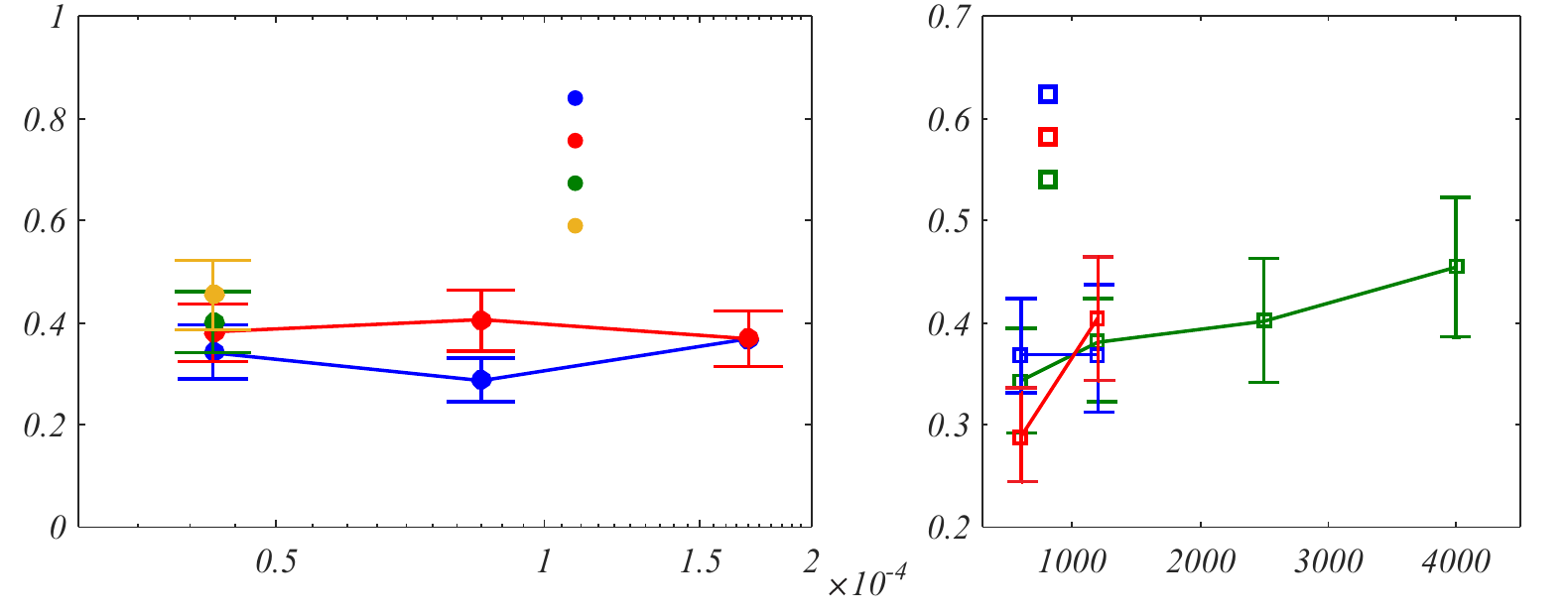}
		\put(-415,110){\rotatebox{-270}{\makebox(0,0)[r]{\strut{} $U_{R}/2\Omega L$}}}%
		\put(-230,148){\makebox(0,0)[r]{\strut{} $Re_Q$}}%
		\put(-231,135){\makebox(0,0)[r]{\strut{} $600$}}%
		\put(-229,123){\makebox(0,0)[r]{\strut{} $1200$}}%
		\put(-229,112){\makebox(0,0)[r]{\strut{} $2500$}}%
		\put(-229,100){\makebox(0,0)[r]{\strut{} $4000$}}%
		\put(-285,0){\makebox(0,0)[r]{\strut{} $Ek$ }}%
		\put(-400,160){\makebox(0,0)[r]{\strut{} $a)$ }}%
		\put(-185,110){\rotatebox{-270}{\makebox(0,0)[r]{\strut{} $U_{R}/2\Omega L$}}}%
		\put(-100,148){\makebox(0,0)[r]{\strut{} $Ek$}}%
		\put(-80,135){\makebox(0,0)[r]{\strut{} $17.0\times10^{-5}$}}%
		\put(-80,123){\makebox(0,0)[r]{\strut{} $8.50\times10^{-5}$}}%
		\put(-80,112){\makebox(0,0)[r]{\strut{} $4.25\times10^{-5}$}}%
		\put(-65,0){\makebox(0,0)[r]{\strut{} $Re_Q$ }}%
		\put(-170,160){\makebox(0,0)[r]{\strut{} $b)$ }}%
	}%
	\caption{Reference velocity $U_{R}$ based on the point of onset of the third advection phase normalized by the inertial wave velocity $2\Omega L$ versus a) $Ek$ and b) $Re_Q$. Only experiments where the onset of third phase was observed were considered.}
	\label{fig8}
\end{figure}
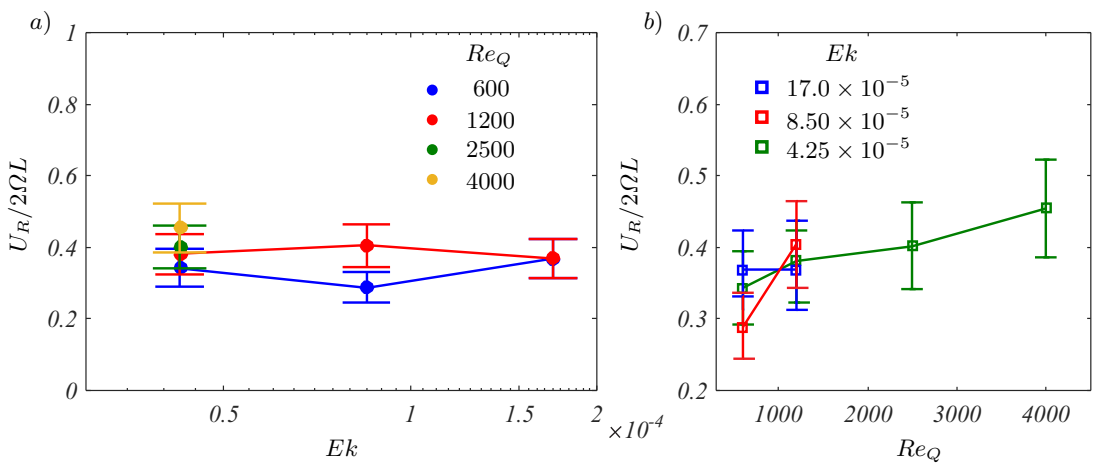

A third phase can be identified in figure \ref{fig7} where the advection speeds up again, \emph{i.e.}, less momentum than in the previous phase is being transported by inertial waves. To understand why inertial waves suddenly loose their efficiency, we calculated the velocity $U_{R}$ of a signal that would have been emitted at the onset of the jet, that would have travelled all the way up to the upper boundary of the vessel and back to the height $z_R$ where the transition takes place, at the time 
it takes place $t_R$, \emph{i.e.} $U_{R}= (2H-z_R-z_a(0))/t_R$.
Figure \ref{fig8}(a) shows that $U_{R}$, normalized by the linear inertial wave velocity at scale $L$, $2\Omega L$, is nearly constant around 0.4 for all values of $Ek$, be it for a very small dependence on $Re_Q$, which may be attributed to weak non-linearities (see figure \ref{fig8}(b)). Hence, the onset of the third phase coincides with the time at which inertial waves have reflected on the top wall and propagated back to the point of transition. This suggests that the loss of momentum transport may result from interferences between inertial waves propagating in opposite directions.
Indeed, if an upward and a downward inertial wave interfere, the upward transport of angular momentum incurred as the upward wave progresses into the still fluid is partially cancelled by the downward momentum transport associated to the reflected wave. Interestingly, the wave velocity associated with the reflected wave is comparatively slower than the fastest upward propagating wave. The reason may be that momentum is not transported by a single wave but by a range of waves of  different lengthscales and velocities. For the momentum transport to drop significantly, a sufficiently broad bandwidth of these waves must have reflected on the top wall, including slower waves, associated to smaller lengthscales.\\

On the subject of wave interactions, it is noteworthy that while interactions between incident waves and waves reflected on the walls of the vessels can sometimes be seen in the patterns, these are indicative of linear waves interference, and not of non-linear wave interaction. As such, the intensity of the waves generated in our experimental setup may be too low for inertial waves to enter the sort of non-linear regime observed when intentionally focusing inertial waves in a region of interaction \citep{duranmatute2013_pre}.\\

Following the suppression of momentum transport by inertial waves, the purely 
advected position resumes its progression at the non-rotating advective pace. 
Remarkably, not only is the velocity but also the position $z^a$ independent of the rotation in this phase, as all positions follow a law:
\begin{equation}
\frac{z^{a}}{L}=(0.48\pm 0.03)\left(\frac{tU_0}{L}\right)^{0.381\pm 0.012}.
\label{eq:za_exp}
\end{equation}
The value of the exponent, lower than the 0.5 value expected for pure advection may reflect that propagation by inertial waves isn't entirely cancelled, as the reflected waves are less intense than the incident ones. Importantly, the dynamics observed in the second phase establishes that not only does rotation introduce an additional transport mechanism with inertial waves, but advection itself is suppressed as a result.
Furthermore, the dynamics of the third phase suggest that momentum transport by inertial waves may not be efficient in confined flows, in particular quasi-two dimensional ones, because waves reflecting on the 
boundaries.\\
\section{Transition to inertial wave propagation} 
\label{sec:propagation}
\subsection{Spectral profile of the turbulent front}
\label{sec:spectralprofile}
\begin{figure}
	\centerline{
		\includegraphics{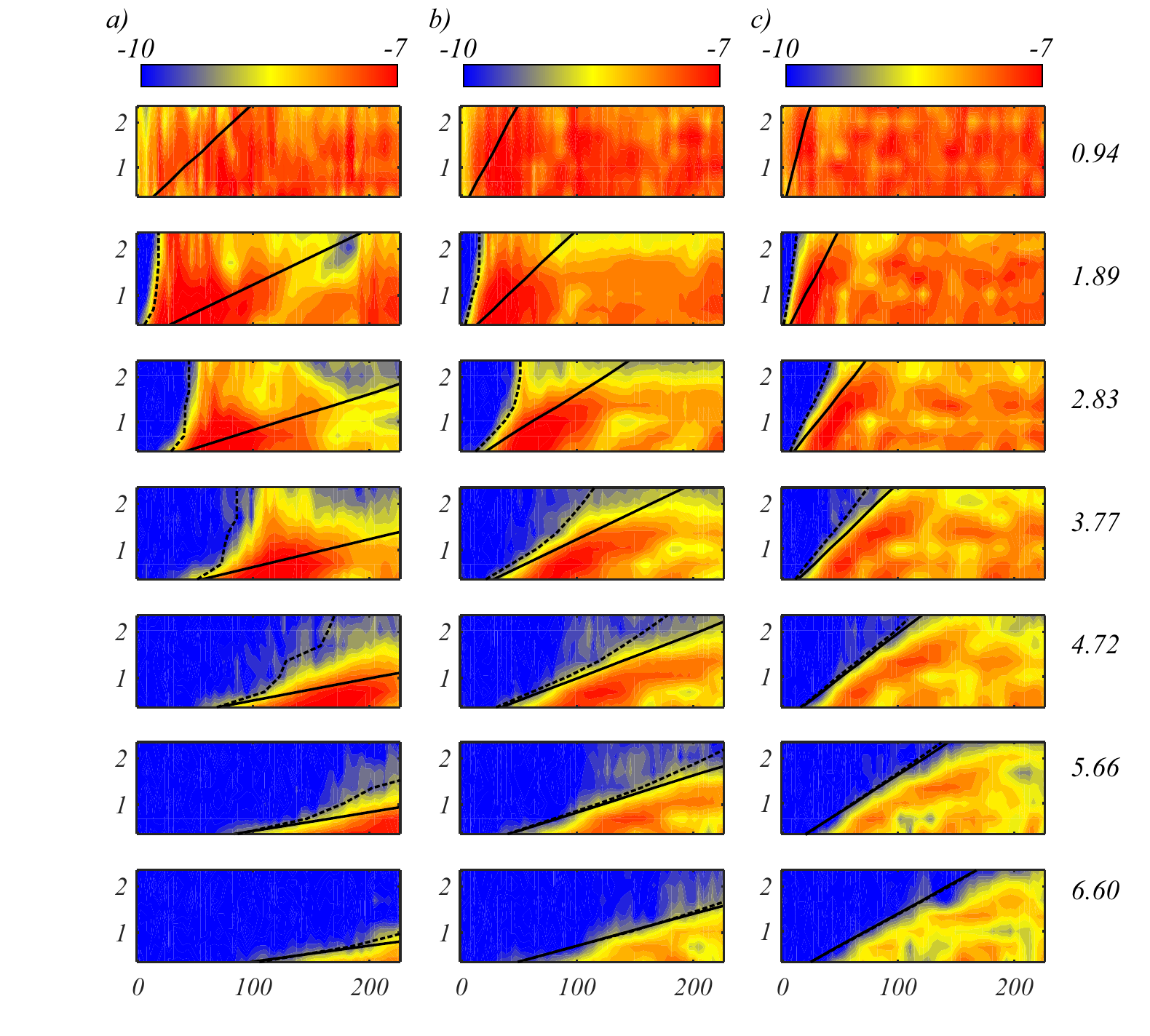}
		\put(-70,395){\makebox(0,0)[r]{\strut{} $\log_{10}(E/U_0^2)$}}%
		\put(-195,395){\makebox(0,0)[r]{\strut{} $\log_{10}(E/U_0^2)$}}%
		\put(-325,395){\makebox(0,0)[r]{\strut{} $\log_{10}(E/U_0^2)$}}%
		\put(-430,205){\rotatebox{-270}{\makebox(0,0)[r]{\strut{}\Large $\frac{Lk}{2\pi}$}}}%
		\put(-13,365){\makebox(0,0)[r]{\strut{} \large $z/L$}}%
		\put(-80,5){\makebox(0,0)[r]{\strut{} $tU_0/L$}}%
		\put(-208,5){\makebox(0,0)[r]{\strut{} $tU_0/L$}}%
		\put(-335,5){\makebox(0,0)[r]{\strut{} $tU_0/L$}}%
	}%
	\caption{Contour plots of $E(k,t)$ across a number of heights $z/L$ for $Re_Q=1200$ at a) $Ek=17.0\times10^{-5}$, b) $Ek=8.50\times10^{-5}$ and c) $Ek=4.25\times10^{-5}$. The solid black line represents the shape of the energy contours assuming propagation is fully driven by inertial waves, \emph{i.e.} $\tau=z/v_g(k)$. Dashed black lines represent the the position of a numerical particle, based on (\ref{eq_num_part}).}
	\label{fig9}
\end{figure}
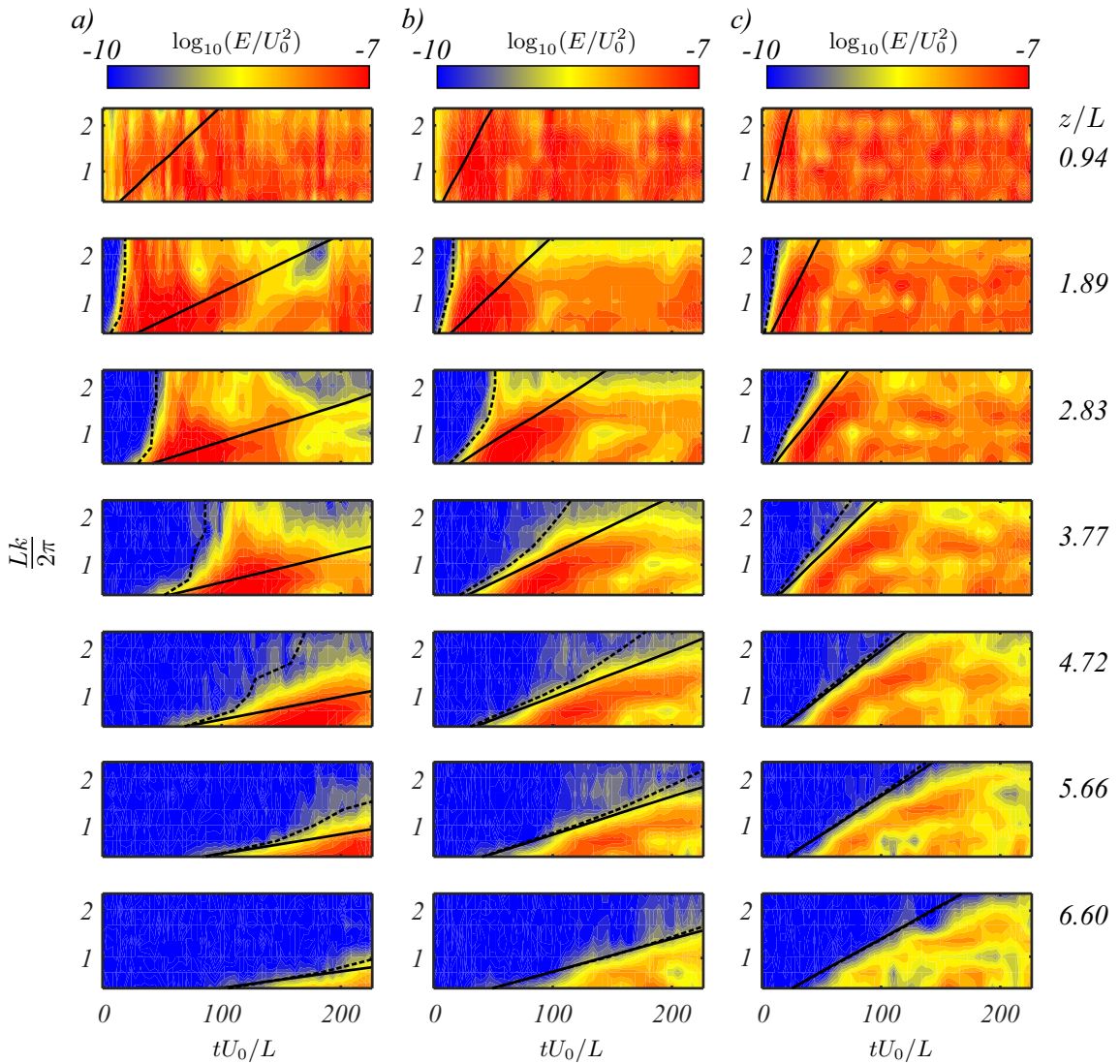

We now seek to characterise the motion of the actual turbulent front in cases where the experiment is rotating, having confirmed that it cannot be explained by advection alone. Figure \ref{fig9} shows the spectral energy density contours of $E(k,z,t)$ at various heights $z$ for $Re_Q = 1200$ across all values of $Ek$ explored. This figure is representative of cases studied for all values of $Re_Q$.
At $z/L=0.94$ there is no discernible difference on the shape of contours between the cases with different values of $Ek$ we investigated. Their near-vertical shape shows that all modes $k$ arrive at the same time and thus all modes progress at approximately the same velocity.\\
For a given value of $Ek$, the spectral contour of the turbulent front progressively changes shape at greater distance $z$ from the bottom wall, exhibiting three regions: the lower wavenumbers arrive at a time indicating that they progress at the group velocity of an inertial wave of the same wavenumber (marked by solid lines). At the higher wavenumbers, by contrast, the front continues to exhibit the flat profile that characterises advection by the jet. These two regions of the front are linked up by a rather narrow transition region. As $z$ increases, the low-wavelength region occupies an increasingly large part of the spectrum, while the high-$k$ advective region shrinks and eventually disappears in all cases we investigated. This is consistent with the morphology of the jet which spreads and therefore slows down away from the source, implying that advection progressively weakens as $z$ increases. For higher rotation (lower values of $Ek$) pictured on the different columns of figure \ref{fig9}, the transition between the propagative and the advective parts of the front becomes increasingly sharp and displaces towards increasingly higher wavenumbers.\\ 
The overall picture is that structures of higher wavenumbers are advected by the jet whereas at low wavenumbers, larger structures propagate with inertial waves. As the  Coriolis force that underpins inertial waves progressively overruns inertial forces associated to advection (either as $z$ increases or as $Ek$ decreases), low wavenumber propagation invades an increasingly wider waveband at the expense of high-wavenumber advection.\\
\subsection{Transport of individual modes}
A finer perspective on the mechanism at play can be gained by tracking individual modes as they are transported along the jet. Considering individual modes offers the opportunity to compare their propagation to the group velocity of inertial waves of the same wavevector along their trajectory.
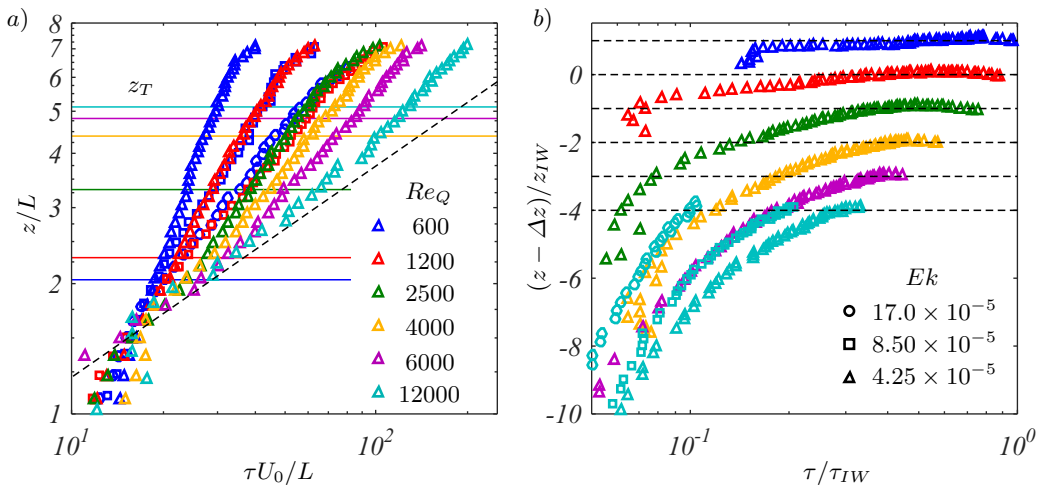
\begin{figure}
	\centerline{
		\includegraphics{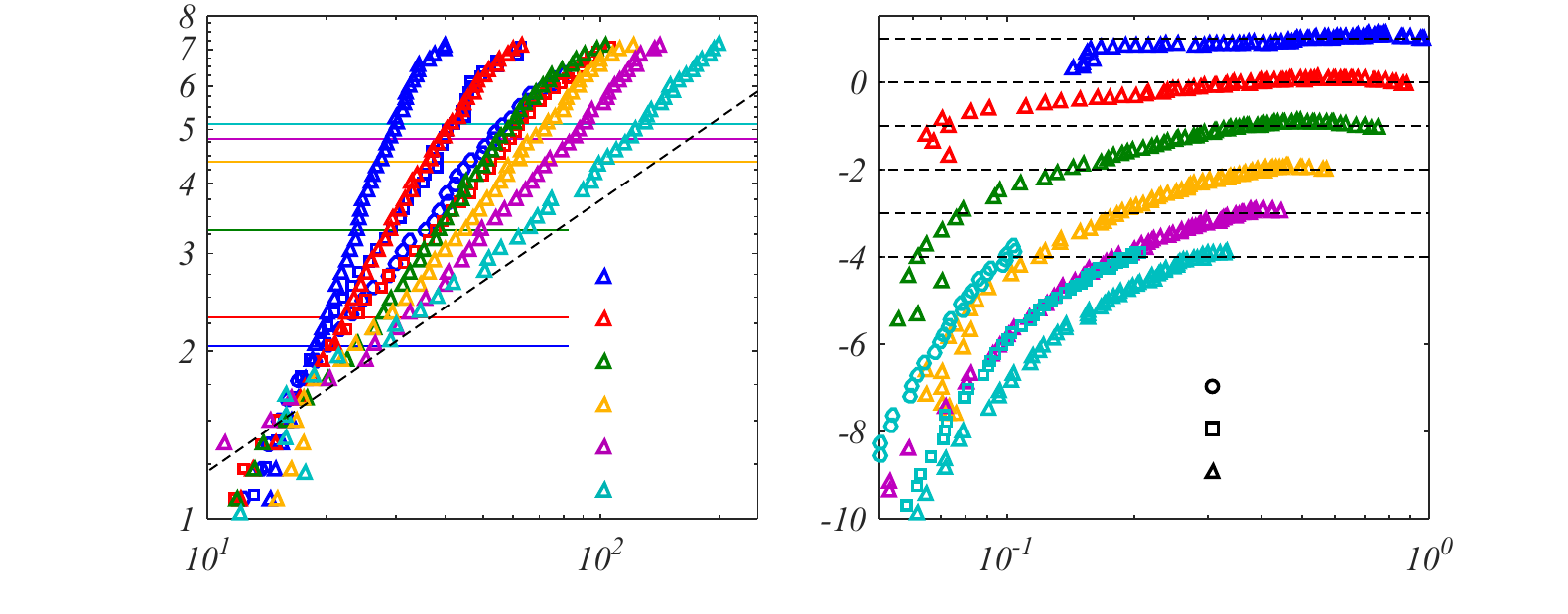}
		\put(-305,5){\makebox(0,0)[r]{\strut{} $\tau U_0/L$}}%
		\put(-412,110){\rotatebox{-270}{\makebox(0,0)[r]{\strut{} $z/L$}}}%
		\put(-95,5){\makebox(0,0)[r]{\strut{} $\tau/\tau_{IW}$}}%
		\put(-220,130){\rotatebox{-270}{\makebox(0,0)[r]{\strut{} $(z-\Delta z)/z_{IW}$}}}%
		\put(-365,150){\makebox(0,0)[r]{\strut{} $z_T$}}%
		\put(-412,175){\makebox(0,0)[r]{\strut{} $a)$}}%
		\put(-215,175){\makebox(0,0)[r]{\strut{} $b)$}}%
		\put(-253,110){\makebox(0,0)[r]{\strut{} $Re_Q$}}%
		\put(-254,97){\makebox(0,0)[r]{\strut{} $600$}}%
		\put(-252,84){\makebox(0,0)[r]{\strut{} $1200$}}%
		\put(-252,72){\makebox(0,0)[r]{\strut{} $2500$}}%
		\put(-252,59){\makebox(0,0)[r]{\strut{} $4000$}}%
		\put(-252,46){\makebox(0,0)[r]{\strut{} $6000$}}%
		\put(-250,34){\makebox(0,0)[r]{\strut{} $12000$}}%
		\put(-70,78){\makebox(0,0)[r]{\strut{} $Ek$}}%
		\put(-48,65){\makebox(0,0)[r]{\strut{} $17.0\times10^{-5}$}}%
		\put(-48,53){\makebox(0,0)[r]{\strut{} $8.50\times10^{-5}$}}%
		\put(-48,41){\makebox(0,0)[r]{\strut{} $4.25\times10^{-5}$}}%
	}%
	\caption{a) Arrival time $\tau$ at height $z$ for mode $k_{1}$ at $Ek=4.25\times10^{-5}$ (triangles), $Ek=8.50\times10^{-5}$ (squares) and $Ek=1.70\times10^{-4}$ (circles). The dashed line represents (\ref{eq_Z_no_rot}). Coloured lines mark $z_T$ where the motion of the turbulent front has transitioned to the propagative mechanism  at $Ek=4.25\times10^{-5}$. b) $z$ and $\tau$ normalized by propagation of inertial wave with wavenumber $k_1$, represented by the dashed lines. Position $z$ is shifted by $\Delta z=z_{IW}(t)-z(t)$. Datasets have been further shifted down by a constant of 1 with respect to one another for clarity.}
	\label{fig10}
\end{figure}
Figure \ref{fig10} (a) shows such trajectories $z(t)$ for mode $k_{1}$, for several values of $Re_Q$ at $Ek=4.25\times10^{-5}$. The dashed line shows the trajectory of the turbulent front when $Ek=\infty$, \emph{i.e.} driven by advection only. Trajectories at all $Re_Q$ initially follow the advection trajectory and separate at a height which increases with $Re_Q$. Past this point, mode $k_1$ progresses faster than if it was advected.\\
To highlight regions of the trajectory that are governed by inertial waves propagation, the trajectories of mode $k_1$ are plotted in figure  \ref{fig10} (b) for several values of $Ek$, using variables $(z-\Delta z)/z_{IW}$ and $\tau/\tau_{IW}$, where $z_{IW}(t)=2\Omega t/k$, $z_{IW}(\tau_{IW})=H$ and $\Delta z$ is the offset between $z$ and $z_{IW}$, measured near the top of the tank.
In these new variables, displacements at the group velocity of mode $k_1$ follow horizontal lines. As expected, trajectories start away from the horizontal propagation lines in the initial advective phase identified in figure \ref{fig10}(a), but gradually bend toward them to end up following them closely. This shows that inertial wave propagation eventually takes over advection. For $Re_Q=12000$ and $Ek=17.0\times10^{-5}$ trajectories barely meet the theoretical propagation line, indicating that propagation never fully takes over within our experimentally accessible parameters. Overall convergence is all the faster as $Re_Q$ and $Ek$ are low, as inertial forces delay the transition from advection to propagation, while rotation accelerates it.\\
To quantify the transition from the advective to the propagative mechanism, we define the point of transition as $z_T=|z/(z_{IW}+\Delta z)-1|\leq\beta$, where $\beta$ is a chosen threshold value. 
The value of $\beta$ has to be chosen as low as possible, however as $\beta$ is lowered the results become increasingly susceptible to experimental noise. To keep noise to a low level, we chose $\beta=$ 0.2 and verified that the results were independent of the exact value we chose.
Figure \ref{fig11} shows $z_T$ for $k_1$ across all $Re_Q$ and $Ek$ explored, with the exception of those  where the transition was not fully achieved (such as for $Re_Q=12000$ and $Ek=17.0\times10^{-5}$). Values of $z_T$ mostly obey a scaling dependent on the Rossby number only:
\begin{equation}
z_T/L\simeq (8.96\pm 0.74) Ro_Q^{1/2}.
\label{eq:zT}
\end{equation}
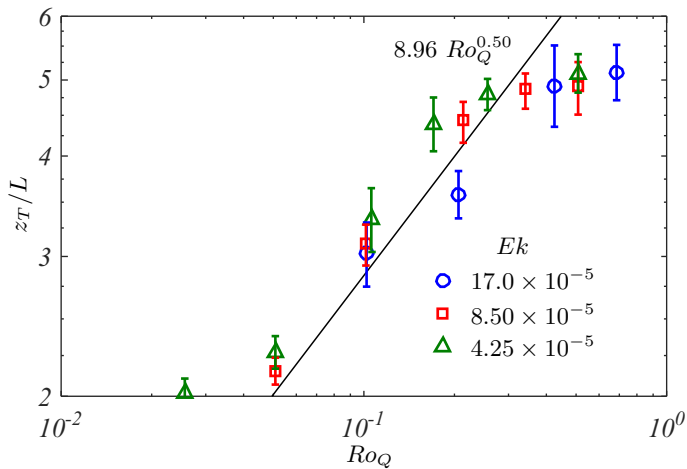
\begin{figure}
	\centerline{
		\includegraphics{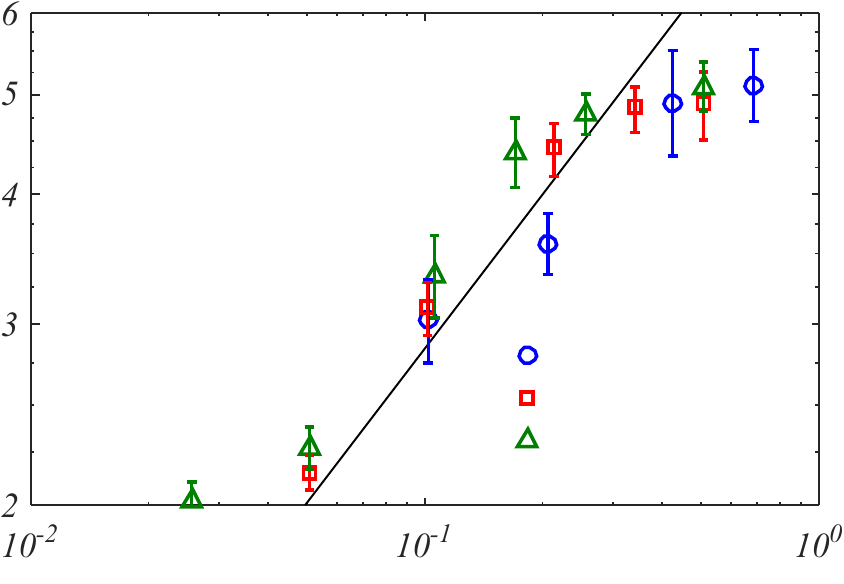}
		\put(-110,0){\makebox(0,0)[r]{\strut{} $Ro_Q$}}%
		\put(-250,105){\rotatebox{-270}{\makebox(0,0)[r]{\strut{} $z_T/L$}}}%
		\put(-65,152){\makebox(0,0)[r]{\strut{} $8.96~Ro_Q^{0.50}$}}%
		\put(-34,65){\makebox(0,0)[r]{\strut{} $17.0\times10^{-5}$}}%
		\put(-59,78){\makebox(0,0)[r]{\strut{} $Ek$}}%
		\put(-34,52){\makebox(0,0)[r]{\strut{} $8.50\times10^{-5}$}}%
		\put(-34,40){\makebox(0,0)[r]{\strut{} $4.25\times10^{-5}$}}%
	}
	\caption{Height $z_T$ beyond which the displacement of scales of wavenumber $k_1$ are driven by the propagative mechanism.}
	\label{fig11}
\end{figure}
A few points depart from this law for $Ro_Q>3\times10^{-1}$. We could verify that this behaviour is an artefact of the method used to determine $z_T$, as lowering the value of $\beta$ shifts this point to higher values of $Ro_Q$ and $z_T/L$. 
Scaling (\ref{eq:zT}) can be understood by considering that at the transition between the two phases, the length of the jet  $z_T$ has reached a point where Coriolis forces are sufficient to balance inertia. 
Considering $z_T$ as the largest lengthscale, in dimensional terms, it must satisfy $U(z_T)/z_T\sim2\Omega$. In the absence of rotation effects, the jet develops as $U(z)/U_0\sim d/z$ \citep[p.100]{pope}, so $z_T$ must scale as $z_T\sim(U_0d/2\Omega)^{1/2}$, or equivalently, $z_T/L\sim Ro_Q^{1/2}$, as in (\ref{eq:zT}). A similar criterion was put forward by \cite{burmann2018_pf} to explain the breakdown of inertial wave propagation in a spun up cylinder where the waves were emitted by a topography of the bottom wall.
When turbulence is forced by an oscillating grid, \citet{dickinson83} similarly observe that the progression of the front is not affected by rotation in the early stages up to a critical distance, which these authors express (in our notations) as $z_T\simeq 0.36(fS^2/\Omega)^{1/2}$, in terms of the frequency $f$ and stroke $S$ of the grid. As such, $fS$ is equivalent to forcing velocity $U_0$ and the scaling for $z_T$ associated to the oscillating grid can be rewritten $z_T/S\simeq 0.36 Ro_Q^{1/2}$.
It is similar to (\ref{eq:zT}), even though reference lengthscales $S$ and $L$ are not necessarily directly comparable and the upward motion imprinted by the jet may contribute to stretch the patch upwards. It is noteworthy that the transition point for mode $k_1$ coincides with the transition point for the whole front. Indeed, wavenumber $k_1$ corresponds to the largest lengthscale, with the fastest inertial wave within the set of wavelengths captured  within the visualisation area. The fact that no inertial wave travels faster than that of scale $k_1$ confirms that $W\simeq 3L$ captures the largest scales of the turbulent patch. As such the visualisation area we chose is suitable to capture the propagative processes responsible for momentum transport.
For the same reason, the point  $z_T$ also corresponds to the point of transition where advection itself starts being suppressed by the effects of rotation (see section \ref{sec:advection}).\\
\subsection{Scaling for the transition between advection and propagation}
The example of $k_1$ illustrates that fluctuations are first advected in the low part of the jet, as advection dominates near the injection point. As they progress through the fluid domain, advection subsides as the jet spreads. At the same time, the mean centreline velocity decreases and propagation by inertial waves takes over as the main transport mechanism. The last step is to understand how this mechanism expresses at other wavelengths $k>k_1$.
To this end, we first note from figure \ref{fig10} and \ref{fig11} that all curves for the displacement of fluctuations of wavenumber $k_1$ gradually transition away from the pure advection trajectory and converge to the propagative trajectory at $z=z_T$. At this point their displacement velocity matches to the propagation velocity of linear inertial waves. Expressing this property for fluctuations of wavelength $k$ yields the condition (here dimensionally written)
\begin{equation}
U(z)\simeq V_g(k)=\frac{2\Omega}{k}.
\label{eq_U=Uiw}
\end{equation}
In other words, the transition from advection to propagation for fluctuations of wavelength $k$ takes place when the local, scale-dependent Rossby number $Ro(k,z)=kU(z)/2\Omega$ reaches unity. Another way to express this is that fluctuations are advected at the fastest of the local advection velocity and the group velocity of inertial waves.\\
To test this criterion on the entire spectrum, we calculate the arrival time of fluctuations for $k\in[0,40]$, for the values of $z$ displayed on figure \ref{fig9}, using the modified expression of the Lagrangian flow:
\begin{equation}
z(k,t)=\int_0^t {\rm max} \left\{\max_x \{\mathbf u(x,z(t'),t'))\cdot\mathbf e_z\},v_g(k)\right\} dt'.
\label{eq_num_part}
\end{equation}
From this expression, we extract the arrival time $\tau(z,k)$ of fluctuations with wavenumber $k$ at height $z$, which forms the spectral shape of the turbulent front. The results are reported on figure \ref{fig9}, which is representative of all other values of $Re_Q$ we considered. In all cases, the motion of a numerical particle subject to (\ref{eq_num_part}) matches the actual contours of $E(k,z,t)$ closely for $z/L\geq1.5$. It indeed captures all three regions identified in section \ref{sec:spectralprofile}. This indicates in particular, that in the intermediate region, the arrival time results from an initial advective phase of comparable duration to a second propagative phase, so that the arrival time falls somewhere between a pure advective and a pure propagative time.
\section{Conclusion and Discussion}
\label{sec:conclusion}
We have analysed the scale-by-scale transport mechanisms in rotating turbulence. The results were obtained by examining the motion of the turbulent front generated during the transient flow of four jets penetrating into or extracted from a rotating vessel of quiescent fluid, and directed along the axis of rotation. 
In the absence of rotation, the distance from the jet source covered by disturbances evolves (in dimensional variables) as $(z(t)-z_0)/L\simeq 0.377 (U_0t/L)^{0.483}$ ($U_0$ and $L$ are the jet inlet velocity and the distance between the jets respectively). This law  is in good agreement with \citet{long1978_pf}'s law for the global displacement of a turbulent front, with an offset $z_0\simeq0.5-2.0$ cm, incidentally consistent with the values experimentally found by \citet{dickinson1978_pf} in experiments with an oscillating grid. Additionally, we established that this law is valid at all scales, regardless of their transversal wavenumber $k$, and of the Reynolds number based on the inlet jet velocity $Re_Q$. In the presence of rotation, the turbulent front is advected exactly as in the non-rotating case up to a distance $z_T/L\simeq 8.96 Ro_Q^{1/2}$, where the Coriolis force becomes larger than inertia. Past this point, the development of the jet is dominated by the faster propagation of inertial waves. However, since momentum is 
redistributed over a larger volume by inertial waves, it is locally weaker. As a consequence, advection itself is suppressed by rotation.\\
In the last phase of the jet's evolution, inertial waves reflected on the vessel's wall of the fluid vessel interfered with inertial waves travelling up, resulting in a suppression of the total transport by inertial waves. This suggests that in confined flows, inertial waves may not be able to transport momentum 
efficiently. This is particularly relevant in the quasi-two dimensional limit, where our recent 
experiments showed that they were indeed not driving the dynamics \citep{bpt2019_prl}.\\
The scale-by-scale analysis of the propagation enabled us to answer the questions set out in the introduction:
\begin{enumerate}
	\item A clear separation exists between scales advected by inertial waves and by the local mean flow. 
	\item The border between the two regimes is set by the Rossby number based on the transversal wavelength 
	of the  scale considered and the local large scale velocity as $Ro_k(k,\mathbf x )=  kU(\mathbf x)/2\Omega=1$. In that sense, this criterion is local both in space, time and scale.  
\end{enumerate}
The implication of this phenomenology is that the transport of turbulent fluctuations as turbulence progresses into the quiescent fluid follows two phases: one purely controlled by local advection for $Ro_k(k,z)>1$ and one purely controlled by the propagation of inertial waves for $Ro_k(k,z)<1$. The spectral locality of the transition complements the recent evidence for its spatial locality found by \citet{mcdermott2019_jfm}.

In other turbulent flows with more complex flow topology, the same phenomenology would imply that structures may be alternately convected by larger structures and propagated by inertial waves.
However, it is worth pointing out that the fact that advection dominates at a given scale does not mean that inertial waves do not exist at that scale. Just like the transversal sweeping of inertial waves in nearly two-dimensional flows \citep{campagne15_pre}. Axial advection of inertial waves could take place in our setup, but would be shadowed if advection was the fastest mechanism. More generally, our result does not exclude the possibility that inertial waves at small scales may be axially or laterally convected by faster advection too. These remarks apply in particular to non-transient turbulent flows.
Indeed, an important feature of the transient problem studied in this paper is the fact that inertial waves are emitted by random fluctuations in a turbulent region where rotation does not dominate. A similar phenomenology may exist in turbulent flows, even when the macroscopic Rossby number remains well below unity, provided random fluctuations also exist at a sufficiently small scale to escape the influence of rotation. Such fluctuations may act as random sources of inertial waves competing with local advection to transport momentum. Unlike in transient problems where the displacement of the turbulent 
front offers a convenient way to track momentum transport, however, the two mechanisms are more difficult to disentangle in statistically steady turbulence, especially if unlike for the jet, momentum is not advected in a preferred direction.\\  
Finally, while the mechanisms found here do not exclude the possibility that non-linear interactions may participate in the build-up of large quasi-two dimensional structures, they illustrate that linear inertial waves govern transport mechanisms at the large scales, as shown by \citet{davidson06}, but they also dominate down to the level of smaller scales as long as the local balance of Coriolis force and advection favours the former. More generally, it is not unusual that turbulence dynamics be controlled at the scale level by linear processes, as illustrated in magnetohydrodynamic turbulence at low magnetic Reynolds number, where the anisotropy of individual scales is controlled by the balance between inertia and momentum diffusion by the Lorentz force \citep{sm82,pk2014_jfm,bpdd2018_prl}. 
Having said this, linear waves themselves can also interact non-linearly and lead to turbulence when they are sufficiently energised, as demonstrated with inertial waves by \cite{duranmatute2013_pre}.
\\
\\
The authors would like to acknowledge support from the Engineering and  Physical Sciences Research Council [grant number GR/N64519/01] for the manufacture of the rotating turntable facility and  B. Teaca for computational resources used in processing experimental data. AP aknowledges support from the Royal Society under the Wolfson Merit Award Scheme (grant WM140032).

\bibliographystyle{jfm}
\bibliography{references_lib}

\end{document}